\documentclass[12pt]{iopart}

\usepackage{iopams}  

\expandafter\let\csname equation*\endcsname\relax
\expandafter\let\csname endequation*\endcsname\relax

\usepackage{epsfig, amsmath, amssymb}
\newtheorem{theorem}{Theorem}
\newtheorem{prop}{Proposition}

\usepackage{hyperref}
\usepackage{lineno}

\usepackage{xcolor}
\usepackage{cite}

\begin{document}

\title[VQC-MLPNet: An Unconventional Hybrid Architecture for QML]{VQC-MLPNet: An Unconventional Hybrid Quantum-Classical Architecture for Scalable and Robust Quantum Machine Learning}

\author{Jun Qi$^{1*}$, Chao-Han Yang$^{2}$, Pin-Yu Chen$^{3*}$, Min-Hsiu Hsieh$^{4*}$}

\address{1. School of Electrical and Computer Engineering, Georgia Institute of Technology, Atlanta, GA 30332, USA     \\ 
2. NVIDIA Research, Santa Clara, CA 95051, USA	 \\
3. IBM Thomas J. Watson Research Center, NY, 10598, USA    \\
4. Hon Hai (Foxconn) Quantum Computing Research Center, Taipei, 114, Taiwan   \\
}
\ead{jqi41@gatech.edu, pin-yu.chen@ibm.com, minhsiuh@gmail.com}
\hspace{25mm}\small{* denotes corresponding authors}
\vspace{10pt}


\begin{abstract}
Variational quantum circuits (VQCs) hold promise for quantum machine learning but face challenges in expressivity, trainability, and noise resilience. We propose VQC-MLPNet, a hybrid architecture where a VQC generates the first-layer weights of a classical multilayer perceptron during training, while inference is performed entirely classically. This design preserves scalability, reduces quantum resource demands, and enables practical deployment. We provide a theoretical analysis based on statistical learning and neural tangent kernel theory, establishing explicit risk bounds and demonstrating improved expressivity and trainability compared to purely quantum or existing hybrid approaches. These theoretical insights demonstrate exponential improvements in representation capacity relative to quantum circuit depth and the number of qubits, providing clear computational advantages over standalone quantum circuits and existing hybrid quantum architectures. Empirical results on diverse datasets, including quantum-dot classification and genomic sequence analysis, show that VQC-MLPNet achieves high accuracy and robustness under realistic noise models, outperforming classical and quantum baselines while using significantly fewer trainable parameters. This work offers a theoretically grounded and noise-resilient pathway for hybrid quantum–classical learning, advancing the frontiers of quantum-enhanced learning for unconventional computing paradigms in the Noisy Intermediate-Scale Quantum era and beyond.
\end{abstract}

%
%
%
%
%

\section{Introduction}
\label{sec1}

Quantum computing has emerged as a promising paradigm capable of solving complex computational problems beyond the reach of classical systems~\cite{arute2019quantum, preskill2018quantum}. In the Noisy Intermediate-Scale Quantum (NISQ) era, characterized by quantum hardware constraints such as noise, decoherence, and limited qubit count~\cite{bharti2022noisy, leymann2020bitter}, hybrid quantum-classical architectures have gained significant attention as practical pathways for leveraging quantum advantages in machine learning applications~\cite{cerezo2022challenges, biamonte2017quantum, benedetti2019parameterized, schuld2019quantum}. Among the core quantum components used in such frameworks, Variational Quantum Circuits (VQCs) have become a foundational approach in quantum machine learning~\cite{cerezo2021variational, benedetti2019parameterized, cong2019quantum, abbas2021power, beer2020training}. VQCs are parameterized quantum models trained via classical optimization loops and are often regarded as quantum analogs of neural networks due to their layered gate structures and tunable parameters~\cite{havlivcek2019supervised, mitarai2018quantum,  schuld2019quantum}. Nevertheless, VQCs encounter critical challenges, including limited representation capacity stemming from their inherent linearity~\cite{du2020expressive}, difficulties with scalable data encoding~\cite{schuld2021effect, huang2021power, caro2022generalization}, optimization problems arising from a non-convex training landscape~\cite{holmes2022connecting, coyle2020born, du2021learnability}, and the well-known barren plateau phenomenon~\cite{mcclean2018barren, holmes2022connecting}.

\begin{figure}
\centerline{\epsfig{figure=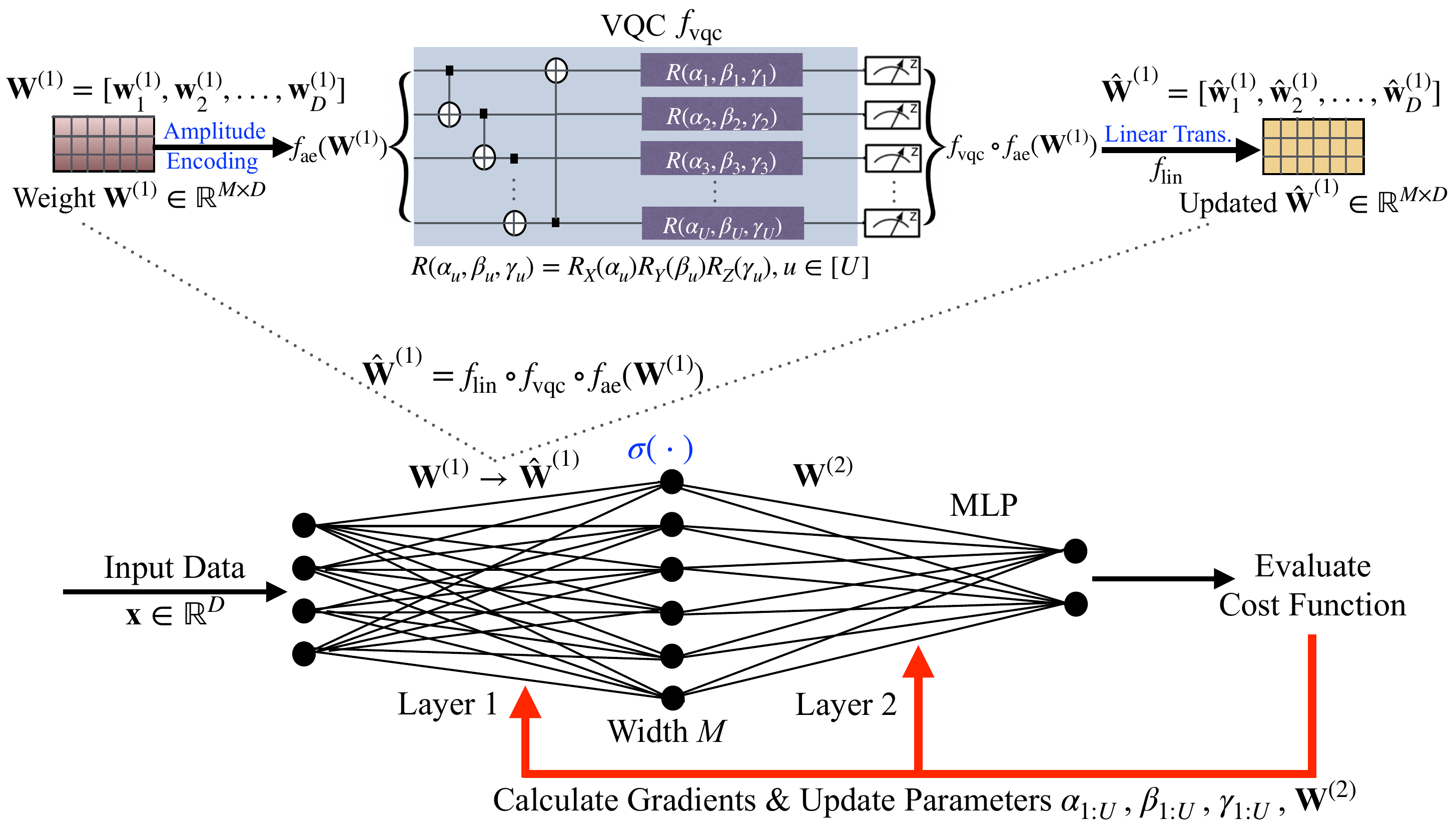, width=125mm}}
\caption{{\it An Illustration of the VQC-MLPNet structure.} The VQC-MLPNet architecture is a hybrid quantum-classical neural network where the VQC generates the MLP parameters $\hat{\textbf{W}}^{(1)}$ for its first hidden layer. During training, the VQC uses amplitude encoding to encode $\textbf{W}^{(1)}$. The transformed parameters $\{\alpha_{1:U}, \beta_{1:U}, \gamma_{1:U} \}$ in $f_{\rm vqc}$ are updated through quantum operations before being integrated into the first hidden layer of the MLP. Once trained, the VQC is no longer needed for inference, making the model scalable for deployment.}
\label{fig:mlp_vqc}
\end{figure}

This work proposes VQC-MLPNet, as shown in Figure~\ref{fig:mlp_vqc}, a hybrid quantum-classical neural network architecture that integrates a Variational Quantum Circuit (VQC) with a classical Multi-Layer Perceptron (MLP)~\cite{hassoun1995fundamentals} to overcome these fundamental limitations. Specifically, our proposed architecture employs a VQC to generate a subset of the MLP's weight parameters, effectively combining quantum-enhanced feature embeddings with the classical neural network's nonlinear expressivity. This quantum-guided parametrization architecture substantially boosts representation power, generalization performance, and optimization stability compared to traditional VQCs. Importantly, VQC-MLPNet maintains computational scalability and practical feasibility for near-term quantum devices by operating classically during inference, making it especially suitable for deployment on current quantum hardware.

Recent hybrid models have sought to address these issues by integrating VQCs with other computational structures. For instance, models like Quantum Convolutional Neural Networks (QCNNs)~\cite{cong2019quantum} have demonstrated strong generalization capabilities by using quantum convolutional and pooling layers to extract hierarchical features, and TTN-VQC~\cite{qi2023theoretical, chen2022quantumCNN} has successfully demonstrated improved expressivity by integrating classical tensor-train networks with variational quantum classifiers. However, like many quantum-centric architectures~\cite{kerenidis2024quantum, di2022dawn}, their practical performance can be constrained by optimization challenges such as barren plateaus and sensitivity to hardware noise in the NISQ era~\cite{nguyen2024theory, caro2022generalization, cerezo2022challenges}. In contrast, VQC-MLPNet offers an alternative design by seamlessly integrating quantum circuit expressivity into a classical neural structure. This approach distinctively addresses these limitations by leveraging classical nonlinear activation functions to improve representation capability and optimization stability, as evidenced by our favorable NTK properties, providing a robust and scalable solution suited for current quantum computing landscapes.

\begin{figure}
\centerline{\epsfig{figure=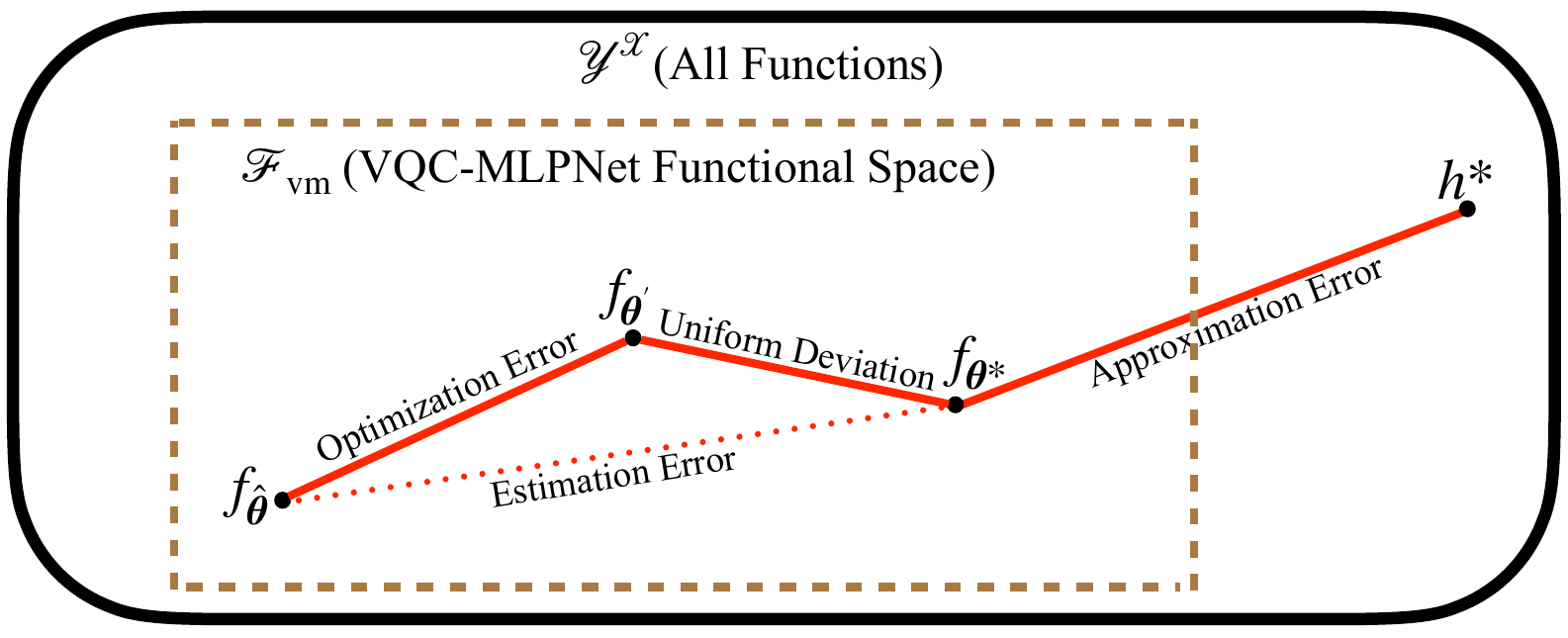, width=105mm}}
\caption{{\it The error performance analysis of VQC-MLPNet illustrates the decomposition of total learning error into three key components}. Given a target function $h^{*}$, the optimal VQC-MLPNet introduces an approximation error relative to $h^{*}$. The empirical risk minimizer $f_{\boldsymbol{\theta}'}$, obtained from training data, results in a uniform deviation from $f_{\boldsymbol{\theta}^{*}}$. Finally, the algorithmically returned operator $f_{\boldsymbol{\hat{\theta}}}$, derived through gradient-based optimization, incurs an optimization error concerning $f_{\boldsymbol{\theta}'}$.}
\label{fig:error}
\end{figure}

To rigorously assess the effectiveness of VQC-MLPNet, as shown in Figure~\ref{fig:error}, we perform a structured theoretical framework of risk decomposition~\cite{bach2024learning, mohri2018foundations} that systematically factorizes the total learning error into three key contributing components: approximation error, uniform deviation, and optimization error. As illustrated in Figure~\ref{fig:error}, given a target operator $h^{*}$, we aim to identify the optimal parametric VQC-MLPNet operator $f_{\boldsymbol{\theta}^{*}}$ within the VQC-MLPNet functional space $\mathcal{F}_{\rm vm}$, where $\boldsymbol{\theta}$ represents the VQC-MLPNet parameters in the VQC-MLPNet parameter space. 

The deviation between $h^{*}$ and $f_{\boldsymbol{\theta}^{*}}$ defines the approximation error, which quantifies the model's ability to represent the target function. In practice, empirical risk minimization is performed on the training data, yielding an empirical risk minimizer $f_{\boldsymbol{\theta}'}$, leading to uniform deviation, which is the discrepancy between $f_{\boldsymbol{\theta}'}$ and $f_{\boldsymbol{\theta}^{*}}$. Furthermore, the final VQC-MLPNet operator $f_{\boldsymbol{\hat{\theta}}}$ is obtained using a gradient-based optimization algorithm, introducing an optimization error that measures the difference between $f_{\boldsymbol{\hat{\theta}}}$ and $f_{\boldsymbol{\theta}'}$. Moreover, the uniform deviation and optimization error sum constitute the estimation error, representing the discrepancy between the learned model $f_{\boldsymbol{\hat{\theta}}}$ and the best achievable operator $f_{\boldsymbol{\theta}^{*}}$.  

Mathematically, the framework of risk decomposition considers the excess risk of $f_{\boldsymbol{\hat{\theta}}}$ as $\mathcal{R}(f_{\boldsymbol{\hat{\theta}}}) - \mathcal{R}(h^{*})$, which can be decomposed as the sum of approximation error and estimation error as shown below: 
\begin{equation}
\mathcal{R}(f_{\hat{\boldsymbol{\theta}}}) - \mathcal{R}(h^{*}) = \underbrace{ \left\{ \mathcal{R}(f_{\boldsymbol{\theta}^{*}}) - \mathcal{R}(h^{*}) \right\} }_{\textcolor{blue}{\text{Approximation  Error}}} +  \underbrace{ \{ \mathcal{R}(f_{\hat{\boldsymbol{\theta}}}) - \mathcal{R}(f_{\boldsymbol{\theta}^{*}}) \} }_{\textcolor{blue}{\text{Estimation Error}}},
\end{equation}
where we separately define $\mathcal{R}(\cdot)$ and $\mathcal{\hat{R}}(\cdot)$ as expected risk and empirical risk, and the estimation error $ \mathcal{R}(f_{\hat{\boldsymbol{\theta}}}) - \mathcal{R}(f_{\boldsymbol{\theta}^{*}})$ can be further decomposed as: 
\begin{equation}
\begin{split}
 \mathcal{R}(f_{\hat{\boldsymbol{\theta}}}) - \mathcal{R}(f_{\boldsymbol{\theta}^{*}}) &= \{	\mathcal{R}(f_{\hat{\boldsymbol{\theta}}}) -  \hat{\mathcal{R}}(f_{\hat{\boldsymbol{\theta}}}) \}	+	\{ \hat{\mathcal{R}}(f_{\boldsymbol{\theta'}}) - \mathcal{R}(f_{\boldsymbol{\theta'}}) \} 	   +    \{ \hat{\mathcal{R}}(f_{\hat{\boldsymbol{\theta}}})  -  \hat{\mathcal{R}}(f_{\boldsymbol{\theta}'}) \}  \\
&\le \underbrace{ 2\sup\limits_{\boldsymbol{\theta} \in \Theta} \left\vert \hat{\mathcal{R}}(f_{\boldsymbol{\theta}}) - \mathcal{R}(f_{\boldsymbol{\theta}}) \right\vert }_{\textcolor{blue}{\text{Uniform Deviation}}} +  \underbrace{ \sup\limits_{\boldsymbol{\hat{\theta}} \in \Theta} \left( \hat{\mathcal{R}}(f_{\boldsymbol{\hat{\theta}}})  -  \hat{\mathcal{R}}(f_{\boldsymbol{\theta}'}) \right) }_{\textcolor{blue}{\text{Optimization Error}}}. 
\end{split}
\end{equation}

We aim to provide upper bounds on each error component to highlight VQC-MLPNet's representation and generalization powers, where the approximation error is related to the representation power, and the estimation error (the sum of uniform deviation and optimization error) corresponds to the generalization capability. This work focuses on the classification tasks, and accordingly, we assume both $\mathcal{R}(\cdot)$ and $\mathcal{\hat{R}}(\cdot)$ as the cross-entropy loss function. The related theoretical results are shown in Table~\ref{tab:comp} and summarized below:

\begin{table}[t]\footnotesize
\center
\caption{Summarizing our theoretical results of VQC-MLPNet and comparing them with MLP and TTN-VQC.}
\renewcommand{\arraystretch}{1.3}
\scalebox{0.85}{
\begin{tabular}{|c||c|c|c|}
\hline
\textbf{Error Component}	&	\textbf{VQC-MLPNet}		&	\textbf{MLP}		&	\textbf{TTN-VQC}\cite{qi2023theoretical}		 \\
\hline
\text{Approximation Error}  & $\frac{C_{1}}{\sqrt{M}} + C_{2}e^{-\alpha L} + \frac{C_{3}}{2^{\beta U}}$ (Theorem~\ref{thm:thm1})  & $\frac{C}{\sqrt{M}}$  & $\frac{C_{1}}{\sqrt{M}} + \frac{C_{2}}{\sqrt{L}}$ \\
\hline 
\text{Uniform Deviation}	&  $\frac{2\Lambda \sqrt{L}r}{\sqrt{\vert S\vert}}$ (Theorem~\ref{thm:thm2})  &  $\frac{2\Lambda\Lambda^{'} r}{\sqrt{\vert S\vert}}$ 	&	$\frac{2r}{\sqrt{\vert S\vert}}\left(\sqrt{\sum_{j=1}^{J} \Lambda_{j}^{2}} + \Lambda' \right)$ \\
\hline
\text{Optimization Condition}	&	\text{NTK}	& \text{Over-Param.} + \text{NTK}		& 	\text{$\mu$-PL condition}	\\
\hline
\text{Optimization Error}	& $C_{0} e^{-\lambda_{\rm min}(\mathcal{K}_{\rm vm})t}$ (Eq. (\ref{eq:ntk1}))  &	$C_{0} e^{-\lambda_{\rm min}(\mathcal{K}_{\rm mlp})t}$	&	$\epsilon_{\rm opt}(0) e^{-\mu t}$		\\
\hline
\end{tabular}}
\label{tab:comp}
\end{table}

\begin{itemize}
\item \textbf{Approximation Error}: VQC-MLPNet achieves significantly improved representation power due to its exponential reduction of approximation error concerning quantum circuit depth ($L$) and the number of qubits ($U$). With the over-parameterization setting for MLP, the MLP's hidden-layer width $(M)$ is significantly large so that the term $\frac{C_{1}}{\sqrt{M}}$ is very close to $0$. Thus, the VQC-MLPNet bound scales as $C_{2}e^{-\alpha L} + \frac{C_{3}}{2^{\beta U}}$, demonstrating exponential expressivity advantages over TTN-VQC, which scales polynomially as $\frac{C_{1}}{\sqrt{M}} + \frac{C_{2}}{\sqrt{L}}$. Since classical MLP, which only scales as $\frac{C}{\sqrt{M}}$, VQC-MLPNet exhibits similar representation capability to MLP. 

\vspace{1mm}

\item \textbf{Estimation Error (Uniform Deviation $+$ Optimization Error)}: For the uniform deviation components, VQC-MLPNet exhibits a better-conditioned $\frac{2\Lambda \sqrt{L}r}{\sqrt{\vert S\vert}}$, outperforming TTN-VQC and classical MLP by effectively controlling model complexity through the circuit depth ($L$). Additionally, the optimization error analysis leveraging the Neural  Tangent Kernel (NTK) theory~\cite{liu2022representation, jacot2018neural, bietti2019inductive} shows that VQC-MLPNet provides favorable training stability and exponential convergence, avoiding the stringent $\mu$-Polyak-Łojasiewicz ($\mu$-PL) condition~\cite{karimi2016linear, qi2020analyzing} required by TTN-VQC. In contrast, classical MLP optimization conditions rely heavily on over-parameterization combined with the NTK theory, potentially requiring significantly more parameters. 
\end{itemize}

By providing explicit upper bounds on each error component, we demonstrate the theoretical superiority of VQC-MLPNet over standalone VQCs and other hybrid architectures, such as TTN-VQC. We explicitly evaluate our model under realistic IBM quantum noise conditions, including amplitude and phase damping rates~\cite{temme2017error}, demonstrating substantial robustness and generalization resilience in the NISQ era. Our empirical validation spans highly interdisciplinary scientific applications: (1) semiconductor quantum-dot charge-state classification~\cite{czischek2021miniaturizing, ziegler2023tuning}, which supports quantum device calibration and optimization relevant to quantum computing hardware, and (2) genome transcription factor binding-site prediction~\cite{tompa2005assessing}, an essential task in computational biology and genomics. Furthermore, our method inherently generalizes to other scientific domains such as quantum chemistry and computational materials science, where scalable, accurate quantum-enhanced machine learning can significantly accelerate discovery and innovation.

\section{Results}
\label{res}

\subsection{VQC-MLPNet Framework}

Our proposed framework, VQC-MLPNet, fundamentally builds up an unconventional hybrid quantum-classical neural network integrating a VQC into the classical MLP pipeline. As shown in Figure~\ref{fig:error}, the core innovation is delegating a subset of the MLP's parameter generation to the VQC, thereby effectively embedding quantum-enhanced features into a classical neural structure. The entire operational flow is as follows: 

\begin{enumerate}
\item \textbf{Parameter Generation}: During training, the VQC $f_{\rm vqc}$ transforms the classical weight $\textbf{W}^{(1)}$ from the MLP into quantum states through amplitude encoding $f_{\rm ae}$. Then, a quantum-enhanced weight matrix $\hat{\textbf{W}}^{(1)}$ is yielded through quantum measurement and a linear transformation $f_{\rm lin}$. 
\begin{equation}
\label{eq:vm}
\hat{\textbf{W}}^{(1)} = f_{\rm lin} \circ f_{\rm vqc} \circ f_{\rm ae}(\textbf{W}^{(1)}). 
\end{equation}

The quantum-enhanced weight $\hat{\textbf{W}}^{(1)}$ then passes through a nonlinear activation function $\sigma(\cdot)$, such as ReLU, followed by another classical linear transformation parameterized by $\textbf{W}^{(2)}$, producing the final output. The proposed VQC-MLPNet architecture simultaneously leverages quantum expressivity and classical representation power by combining quantum-generated parameters with classical nonlinearities. 

\item \textbf{Classical Forward Pass}: The VQC-generated parameters are injected into the weight matrix of the MLP's first hidden layer, which functions exclusively in inference mode to evaluate the loss. The classical MLP acts as a fixed forward pass evaluator, where no parameters within the VQC are trained. In contrast, gradient computations and all parameter updates occur via standard backpropagation. 

\end{enumerate}

The VQC-MLPNet architecture is designed to bridge the gap between purely quantum and classical models. By enabling the VQC to influence the MLP weight space, the model benefits from quantum-enhanced transformations without entirely relying on quantum computation during inference.  As a result, VQC-MLPNet achieves a balanced integration of quantum expressivity and classical trainability, making it highly suitable for practical quantum-enhanced machine learning in the current NISQ era.

\subsection{Variational Quantum Circuits} 

\begin{figure}
\centerline{\epsfig{figure=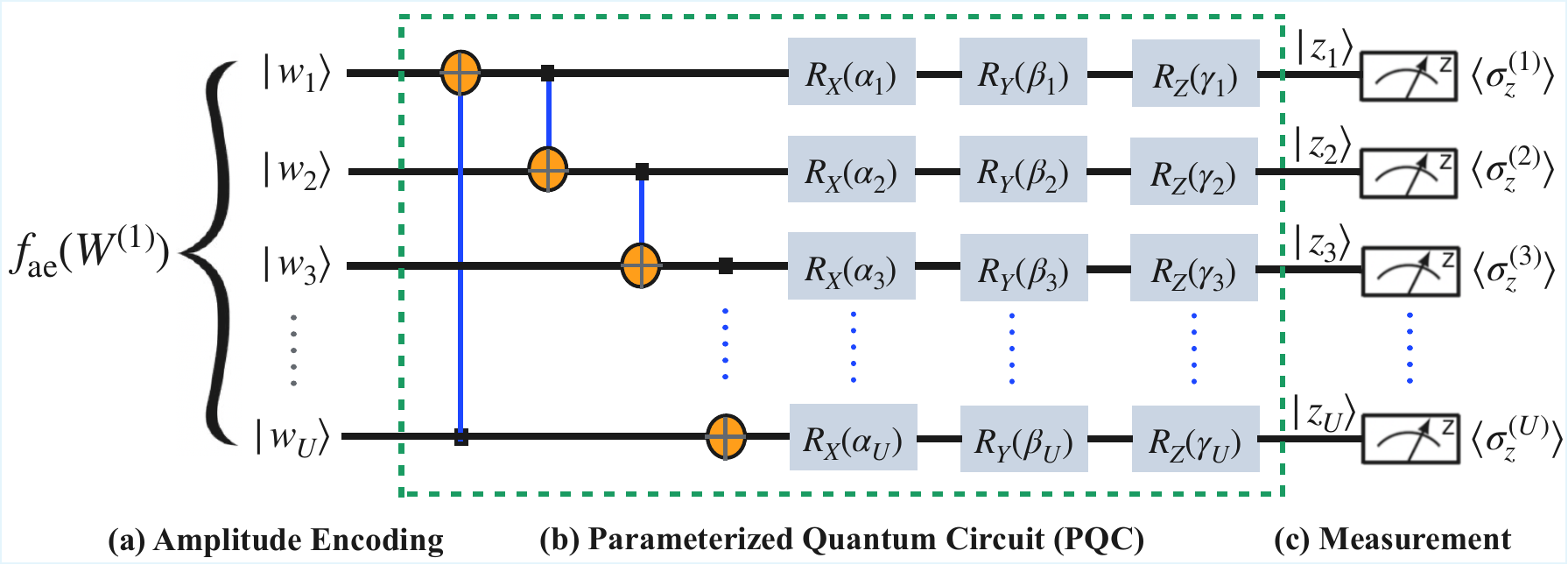, width=135mm}}
\caption{{\it The VQC module in the VQC-MLPNet pipeline}. Each input weight vector from the MLP's first hidden layer is amplitude-encoded into a quantum state over $U$ qubits. The circuit applies parameterized single-qubit rotations $R_{X}(\alpha_u)$, $R_{Y}(\beta_u)$, $R_{Z}(\gamma_u)$ to over $U$ qubits, optionally followed by entangling layers composed of CNOT gates to capture multi-qubit correlations. The PQC model in the green dashed square is repeatedly copied to build a deeper model. The resulting quantum state is measured through Pauli-Z observables ($\vert z_{1}\rangle$, $\vert z_{2} \rangle$, ..., $\vert z_{U}\rangle$), resulting in the expectation values ($\langle \sigma_{z}^{(1)} \rangle$, $\langle \sigma_{z}^{(2)} \rangle$, ..., $\langle \sigma_{z}^{(U)} \rangle$).}
\label{fig:vqc}
\end{figure}

As shown in Figure~\ref{fig:vqc}, a central component of the VQC-MLPNet pipeline is the VQC module, which serves as a quantum parameter generator for the first hidden layer of the MLP. The VQC consists of three main stages: amplitude encoding, parameterized quantum circuit (PQC), and measurement. 

\begin{enumerate}
\item \textbf{Amplitude Encoding}. Each column vector $\textbf{w}_{d}^{(1)} \in \mathbb{R}^{M}$ from the classical weight matrix $\textbf{W}^{(1)} \in \mathbb{R}^{M \times D}$ is normalized and embedded into a quantum state: 
\begin{equation}
\vert \textbf{w}_{d}^{(1)} \rangle = \frac{1}{\lVert \textbf{w}_{d}^{(1)}\rVert_{2}} \sum\limits_{m=0}^{M-1} w_{d}^{(1)}(m) \vert m\rangle, 
\end{equation}
requiring $U$ qubits such that $2^{U} \ge M$. This step encodes classical parameters into the Hilbert space of the quantum register. In the naive case, preparing an arbitrary state with $M$ amplitudes requires $\mathcal{O}(M)$ controlled rotations and $\mathcal{O}(M)$ depth, which scales exponentially with the number of qubits $U$ since $M = 2^{U}$. 

\hspace{6mm} However, recent results on efficient amplitude encoding~\cite{nakaji2022approximate} show that this overhead can be significantly reduced by exploiting structure in the input data. If the weight vectors are sparse with only $k \ll M$ nonzero entries, state preparation can be performed in $\mathcal{O}(k\log M)$. Likewise, if the vectors admit a low-rank tensor network representation (e.g., tensor-train format)~\cite{qi2023theoretical, oseledets2011tensor}, the preparation cost reduces from $\mathcal{O}(M)$ to $\mathcal{O}(U\cdot r^2)$, where $r$ is the bond dimension. 

\hspace{6mm} In parallel, the work~\cite{gonzalez2024efficient} demonstrates approximate encoding strategies that achieve polylogarithmic complexity in $M$, trading off a small, controllable fidelity loss for substantial efficient gains. In practice, these structured encodings and factorizations make amplitude encoding scalable and tractable, mitigating what would otherwise be the main computational bottleneck in our VQC-MLPNet framework.


\vspace{1.5mm}
\item \textbf{Parameteric Quantum Circuit}
Once amplitude encoding is performed, the qubits are evolved under a sequence of trainable rotations, 
\begin{equation}
R(\alpha_{u}, \beta_{u}, \gamma_{u}) = R_X(\alpha_{u}) R_Y(\beta_u) R_Z(\gamma_{u}), \hspace{2mm} u\in [U], 
\end{equation}
alongside entangling CNOT gates to capture correlations across qubits. This produces a quantum-enhanced transformation of the input weights. The PQC depth $L$ determines expressivity: increasing $L$ exponentially reduces the approximation error (see Theorem~\ref{thm:thm1}), though at the cost of additional circuit depth. 

\vspace{1.5mm}
\item \textbf{Measurement and Weight Reconstruction}
After the PQC evolution, each qubit is measured in the computational basis or via Pauli-Z expectation values. These measured outcomes ($\langle \sigma_{z}^{(1)} \rangle$, $\langle \sigma_{z}^{(2)} \rangle$, ..., $\langle \sigma_{z}^{(U)} \rangle$) are aggregated and linearly projected to produce the transformed weight matrix $\hat{\textbf{W}}^{(1)}$. This matrix is then integrated into the MLP pipeline, enabling quantum-enhanced parameterization during training. More importantly, once training is complete, inference proceeds entirely classically with the fixed $\hat{\textbf{W}}^{(1)}$, ensuring scalability and efficiency. 
\end{enumerate}

\subsection{Error Performance Analysis}

To rigorously evaluate VQC-MLPNet, we perform a comprehensive theoretical analysis that decomposes the total learning error into three key components: approximation error, uniform deviation, and optimization error. We derive upper bounds for each error type and systematically demonstrate how VQC-MLPNet outperforms standalone MLPs and traditional hybrid architectures, such as TTN-VQC, in terms of expressivity, generalization capability, and trainability. 

\vspace{1.5mm}

\begin{itemize}
\item \textbf{Mathematical Preliminaries}:

\vspace{0.5mm}

We define the VQC-MLPNet's parameters as the vector $\boldsymbol{\theta} = \boldsymbol{\theta}_{\rm vqc} \oplus  \boldsymbol{\theta}_{W^{(2)}}$, where the VQC's parameter $\boldsymbol{\theta}_{\rm vqc} = [\alpha_{1:U} \hspace{1mm} \beta_{1:U} \hspace{1mm} \gamma_{1:U}]^{\top}$ and $\boldsymbol{\theta}_{W^{(2)}} = [\text{vec}(\textbf{W}^{(2)})]^{\top}$. Then, given two constant constraints $\Lambda_{Q}$ and $\Lambda$, we denote the VQC-MLPNet's functional space as $\mathcal{F}_{\rm vm}$ with its parameter sapce $\Theta$ as: 
\begin{equation}
\Theta = \left\{ \boldsymbol{\theta} =	\boldsymbol{\theta}_{\rm vqc} \oplus  \boldsymbol{\theta}_{W^{(2)}} \hspace{1mm} \vert \hspace{1mm} \lVert \boldsymbol{\theta}_{\rm vqc} \rVert_{2} \le \Lambda_{Q}, \hspace{1mm} \lVert \boldsymbol{\theta}_{W^{(2)}} \rVert_{2}\le \Lambda \right\}. 
\end{equation}

\hspace{6mm} Moreover, we assume the parameterized VQC-MLPNet operator $f_{\boldsymbol{\theta}}$ for its parameters $\boldsymbol{\theta} \in \Theta$. In particular, we separately define $\boldsymbol{\theta}^{*}$ and $\boldsymbol{\theta}^{'}$ as: 

\begin{equation}
\boldsymbol{\theta}^{*} := \arg\inf\limits_{\boldsymbol{\theta} \in \Theta} \mathcal{R}(f_{\boldsymbol{\theta}}) \hspace{4mm} \text{and} \hspace{4mm} \boldsymbol{\theta}^{'} := \arg\inf\limits_{\boldsymbol{\theta} \in \Theta} \mathcal{\hat{R}}(f_{\boldsymbol{\theta}}),
\end{equation}
where $\mathcal{R}(\cdot)$ and $\mathcal{\hat{R}}(\cdot)$ refer to the expected risk and empirical risk, respectively. $\boldsymbol{\theta}^{*}$ and $\boldsymbol{\theta}'$ separately denote the optimal VQC-MLPNet operator and the related empirical risk minimizer. \\

\item \textbf{Approximation Error}:
\vspace{0.5mm}

The approximation error quantifies the discrepancy between the target operator $h^{*}$ and the best achievable approximation $f_{\boldsymbol{\theta}^{*}}$ within the VQC-MLPNet hypothesis space, as formally established in Theorem~\ref{thm:thm1}.

\begin{theorem}[Approximation Error Bound]
\label{thm:thm1}
Given a target operator $h^{*}$, let $f_{\boldsymbol{\theta}^{*}}$ denote the optimal VQC-MLPNet operator within the hypothesis class. Then, the approximation error $\epsilon_{\rm app}$ satisfies the following upper bound: 
\begin{equation}
\label{eq:app_error}
\epsilon_{\rm app} = \mathcal{R}(f_{\boldsymbol{\theta}^{*}}) - \mathcal{R}(h^{*}) \le \frac{C_{1}}{\sqrt{M}} + C_{2} e^{-\alpha L} + \frac{C_{3}}{2^{\beta U}},
\end{equation}
where $M$ is the width of the MLP's hidden layer, $L$ is the depth of the VQC, and $U$ denotes the number of qubits. Constants $C_{1}, C_{2}, C_{3}, \alpha > 0$ are problem-dependent, and $0 < \beta \le \frac{1}{2}$ reflects a scaling penalty associated with amplitude encoding. 
\end{theorem}

\hspace{6mm} Theorem~\ref{thm:thm1} highlights that in an over-parameterized scenario (large width $M$)~\cite{su2019learning}, the approximation error predominantly depends on the VQC structure. Thus, VQC-MLPNet attains a substantially tighter error bound than TTN-VQC, owing to the exponential expressivity improvements by increasing quantum circuit depth $L$ and qubit count $U$. 

\vspace{0.5mm}

\hspace{6mm} Theoretically, increasing $U$ and $L$ significantly enhances approximation power. However, setting large values of these parameters may be challenging. To address this, we provide a guideline in Proposition~\ref{prop:prop1} for choosing a moderate circuit depth $L$ to meet a specified error threshold. 

\begin{prop}[Moderate-depth Circuit Selection]
\label{prop:prop1}
For the expressive error term $C_{2} e^{-\alpha L}$, given a constant $\tau >0$, a logarithmic VQC's depth $L \sim \mathcal{O}(\log(1/\tau))$ suffices to ensure an error remains below $\tau$.
\end{prop}

\vspace{1mm}

\item \textbf{Uniform Deviation}: 

This uniform deviation captures the generalization error, representing the difference between the empirical risk minimizer $f_{\boldsymbol{\theta}^{'}}$ obtained from training data and the optimal $f_{\boldsymbol{\theta}^{*}}$, as formulated in Theorem~\ref{thm:thm2}. 

\begin{theorem}[Uniform Deviation Bound]
\label{thm:thm2}
VQC-MLPNet achieves a favorable bound on uniform deviation due to the controlled complexity of its hybrid architecture: 
\begin{equation}
\begin{split}
\epsilon_{\rm dev} = 2 \sup\limits_{\boldsymbol{\theta} \in \Theta}\left\vert \hat{\mathcal{R}}(f_{\boldsymbol{\theta}}) - \mathcal{R}(f_{\boldsymbol{\theta}}) \right\vert \le \frac{2\Lambda \Lambda_{Q} r}{\sqrt{\vert S \vert}},  \hspace{3mm} \text{s.t.,} \hspace{1mm}  \lVert \boldsymbol{\theta}_{W^{(2)}} \rVert_{2} \le \Lambda, \hspace{1mm} \lVert \boldsymbol{\theta}_{\rm vqc}	\rVert_{2} \le \Lambda_{Q}, 
\end{split}
\end{equation}
where $\Lambda$ and $\Lambda_{Q}$ are model norm constraints, $r$ is the input data norm bound, and $\vert S\vert$ is the number of training samples. 
\end{theorem}

\hspace{6mm} Theorem~\ref{thm:thm2} implies that deeper circuits improve expressivity but could increase model complexity and potential generalization challenges. Furthermore, motivated by the statistical behavior of random quantum circuits, where feature norms typically scale as the square root of depth~\cite{preskill2018quantum, cerezo2021variational, abbas2021power}, we assume $\Lambda_{Q} = \mathcal{O}(\sqrt{L})$. This moderate scaling ensures that the circuit depth enhances model expressivity without excessively inflating the uniform deviation term, supporting both approximation and generalization properties of the VQC-MLPNet architecture. Thus, we refine our uniform deviation bound in Proposition~\ref{prop:prop2}: 

\begin{prop}[Refined Uniform Deviation Bound]
\label{prop:prop2}
Given constraints on the VQC-MLPNet parameters and VQC depth of $\sqrt{L}$, the refined uniform deviation bound is:
\begin{equation}
\epsilon_{\rm dev} \le \frac{2\Lambda \sqrt{L} r}{\vert S \vert}. 
\end{equation}
\end{prop}

\vspace{1mm}

\item \textbf{Optimization Error}: 

To analyze the effectiveness of gradient-based training in VQC-MLPNet, we leverage the Neural Tangent Kernel (NTK) theory. The NTK theory describes how the neural network behaves during training by observing how small parameter changes affect predictions. Unlike TTN-VQC, which relies on strong optimization assumptions such as the Polyak-Łojasiewicz (PL) condition~\cite{karimi2016linear, qi2020analyzing}, VQC-MLPNet naturally ensures exponential convergence due to its well-conditioned NTK and its smallest eigenvalue being not too small, as shown in Eq. (\ref{eq:ntk1}).
\begin{equation}
\label{eq:ntk1}
\epsilon_{\rm opt}(t) = \sup\limits_{\boldsymbol{\hat{\theta}} \in \Theta}\left( \mathcal{\hat{R}}(f_{\boldsymbol{\hat{\theta}}}) -   \mathcal{\hat{R}}(f_{\boldsymbol{\theta}^{*}}) \right) \le C_{0} e^{-\lambda_{\rm min}(\mathcal{K}_{\rm vm})t}, 
\end{equation}
where $C_{0}$ is a constant related to first-order gradients of $f_{\boldsymbol{\theta}}$ w.r.t. paramters $\boldsymbol{\theta}$ at initialization,  $\mathcal{K}_{\rm vm}$ denotes the NTK for VQC-MLPNet, and $\lambda_{\min}(\mathcal{K}_{\rm vm})$ represents its smallest eigenvalue. 

\vspace{0.5mm}

\hspace{6mm} Further decomposing the NTK for the hybrid VQC-MLPNet structure, we have:
\begin{equation}
\label{eq:ntk}
\mathcal{K}_{\rm vm} = \mathcal{K}_{\rm vqc} + \mathcal{K}_{W^{(2)}},
\end{equation}
where $\mathcal{K}_{\rm vqc}$ and $\mathcal{K}_{W^{(2)}}$ represent the NTKs for VQC and the classical output layer of MLP, respectively. Because classical neural networks typically possess better-conditioned NTKs~\cite{lee2019wide}, integrating the classical components of MLP substantially enhances the smallest eigenvalue, significantly improving trainability relative to standalone VQC: 
\begin{equation}
\lambda_{\min}(\mathcal{K}_{\rm vm}) \gg \lambda_{\min}(\mathcal{K}_{\rm vqc}). 
\end{equation}

\hspace{6mm} This result confirms the significant improvement in training stability and efficiency offered by VQC-MLPNet compared to standalone VQCs and TTN-VQC.

\vspace{0.5mm}

\item \textbf{Comparison with MLP and TTN-VQC}:

In summary, compared to classical MLPs, which rely heavily on large-scale parameterization, VQC-MLPNet achieves comparable or superior approximation performance with significantly fewer parameters, thanks to its quantum expressivity. Unlike TTN-VQC, which is constrained by qubit overhead and suffers from optimization difficulties, VQC-MLPNet utilizes classical nonlinear activations to enhance VQC's stability significantly. 
\end{itemize}

\subsection{Empirical Results of Quantum Dot Classification}

To support our theoretical findings, we experimentally evaluate the performance of the proposed VQC-MLPNet model on a binary classification task involving charge stability diagrams of single and double quantum dots (QDs)~\cite{gualtieri2025qdsim}. As illustrated in Figure~\ref{fig:dots}, the dataset contains clean (simulated, noiseless) and noisy (practical, noise-contaminated) charge stability diagrams, clearly indicating charge transition lines. Diagrams labeled `$0$' correspond to single QDs, while those labeled `$1$' correspond to double QDs. We aim to apply quantum machine learning methods to classify these diagrams based on their characteristic transition line patterns. Our experiments are implemented based on Torch Quantum~\cite{hanruiwang2022quantumnas}, and the experimental evaluation is structured into two distinct settings:
\begin{enumerate}
\item \textbf{Representation Power Assessment}: We first examine the representation power of VQC-MLPNet using clean, noiseless charge stability diagrams. Both training and test data are drawn from the simulated, noise-free environment depicted in Figure \ref{fig:dots}(a). This setting explicitly evaluates the model’s capability to accurately represent and distinguish the inherent structural differences between single and double QD configurations without interference from external noise. 

\begin{figure}
\centerline{\epsfig{figure=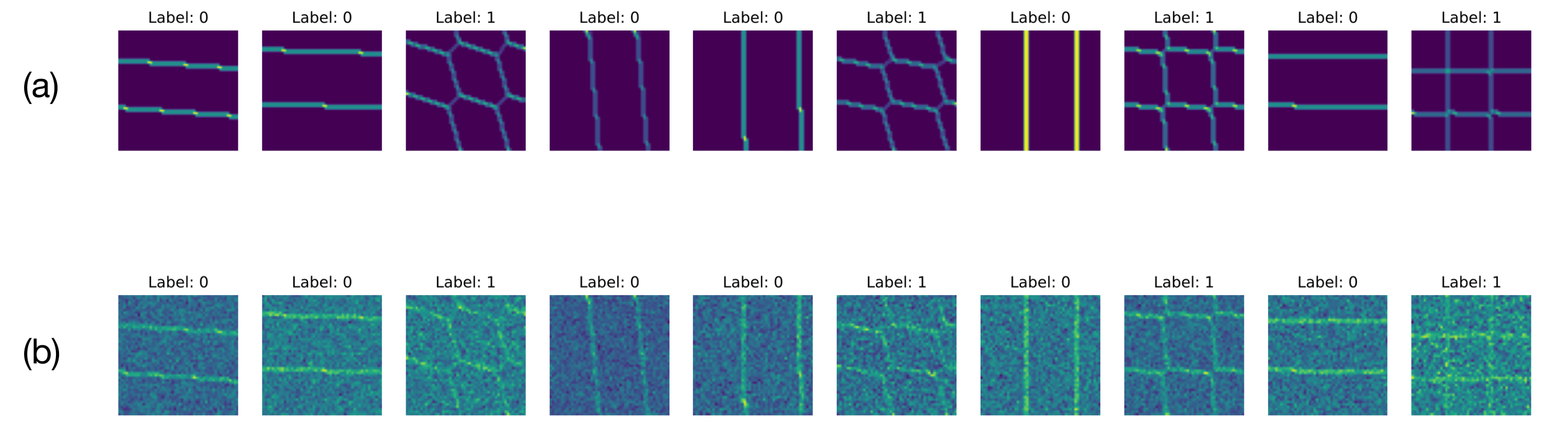, width=150mm}}
\caption{\textbf{Illustration of single and double quantum dot charge stability diagrams used for classification tasks}. (a) Simulated, noise-free charge stability diagrams clearly show charge transition lines. (b) Practical, noise-contaminated stability diagrams reflecting realistic experimental conditions. In both scenarios, Label $0$ represents single QD configurations, and Label $1$ represents double QD configurations. The quantum machine learning model aims to classify these diagrams by identifying characteristic charge transition line patterns, thereby distinguishing between single-dot and double-dot systems. The simulated noise-free diagrams (a) are employed to evaluate the representation power of the models, while the practical, noisy diagrams (b) assess the models' generalization capabilities under realistic noise conditions.}
\label{fig:dots}
\end{figure}

\item \textbf{Generalization Capability Evaluation}: Next, we assess the generalization power of VQC-MLPNet using noisy charge stability diagrams, as shown in Figure \ref{fig:dots}(b), incorporating realistic noise effects encountered in practical quantum experiments. Here, the model trained on noisy data is tested on unseen noisy diagrams, highlighting its robustness and practical utility in realistic quantum environments.

\end{enumerate}

We conducted comparative experiments to evaluate the representation and generalization capabilities of our proposed VQC-MLPNet, the classical MLP, and the previously developed hybrid TTN-VQC architecture. The primary aims of our experimental setup are as follows:

\begin{itemize}
\item To validate that VQC-MLPNet exhibits improved trainability, representation capability, and generalization power relative to MLP and TTN-VQC, consistent with our theoretical analysis.
\item To confirm the theoretical predictions regarding the effects of VQC depth and the number of qubits on the performance of the VQC-MLPNet model.
\item To justify that the quantum-classical VQC-MLPNet architecture is beneficial to the trainability of VQC. 
\item To verify the practical applicability and effectiveness of the VQC-MLPNet framework in a realistic quantum computing environment.
\end{itemize}

The experimental dataset comprises $50 \times 50$ pixel images representing quantum dot charge stability diagrams, including 2,000 simulated noiseless diagrams and 2,000 diagrams generated under realistic experimental noise conditions. We randomly partition the noiseless diagrams into 1,800 training and 200 test samples to evaluate representation capabilities. Similarly, to assess the models' generalization capabilities under realistic noise, we utilize 1,800 noisy diagrams for training and reserve 200 noisy diagrams for testing.

In our experiments, the proposed VQC-MLPNet and the baseline TTN-VQC architectures employ VQCs configured with $20$ qubits and a circuit depth of $6$. All models utilize the cross-entropy loss function, which is optimized through gradient-based parameter updates. For a baseline comparison, we implement an over-parameterized classical MLP with a hidden layer width of $2,048$ (explicitly chosen to maintain the over-parameterized and NTK settings), which results in approximately $5.13$ million parameters, significantly exceeding the number of training data points. In contrast, our proposed VQC-MLPNet architecture substantially reduces the parameter count to approximately 47.46 thousand parameters, achieving a greater than $100$-fold improvement in parameter efficiency due to quantum-enhanced representations.

All models are trained for $20$ epochs using the Adam optimizer with a fixed learning rate of $0.001$. To ensure reproducibility and fair comparison, we conduct $5$ independent runs with different random seeds, where model parameters are initialized from a normal distribution. For quantum-based models, each circuit expectation value is estimated from 4,096 measurement shots, capturing statistical fluctuations due to quantum sampling noise. Quantum circuit simulations are executed on two RTX 4090 GPUs, each equipped with 24 GB of memory.

\begin{itemize}

\item \textbf{Empirical Results on Representation Power}:

\vspace{0.5mm}

\begin{figure}
\centerline{\epsfig{figure=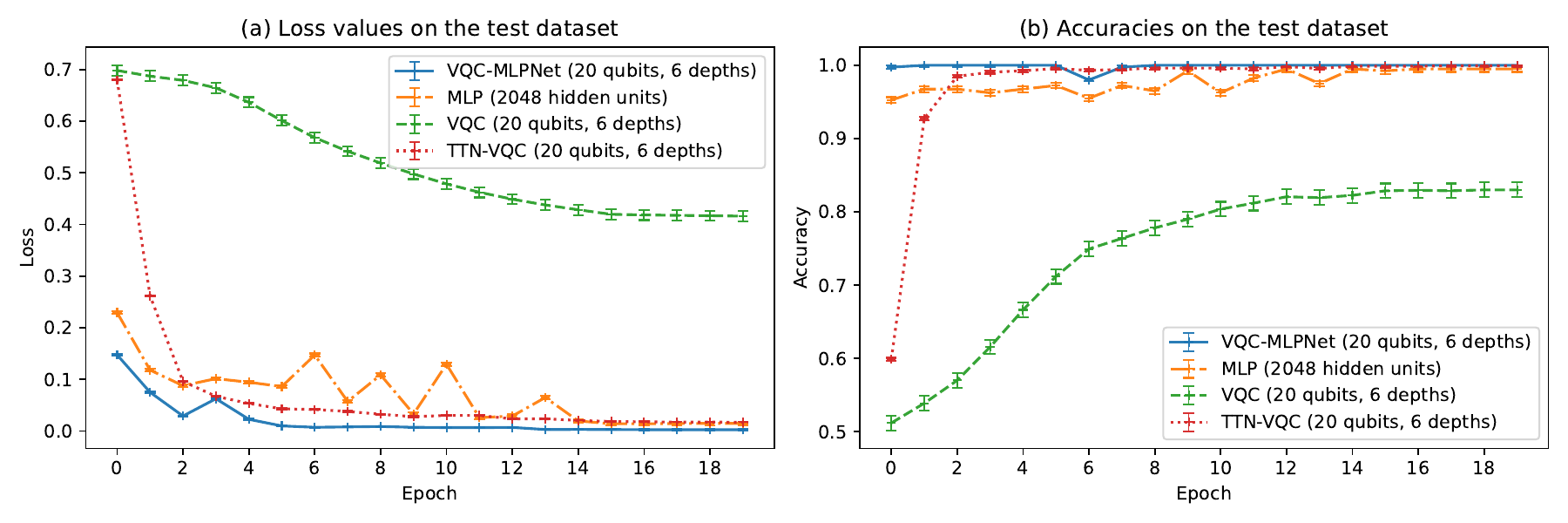, width=160mm}}
\caption{\textbf{Empirical results of quantum dot classification on the test dataset to evaluate models' representation power}.  (a) Test loss and (b) test accuracy over 20 epochs for VQC-MLPNet ($20$ qubits, $6$ layers), TTN-VQC ($20$ qubits, $6$ layers), a plain VQC ($20$ qubits, $6$ layers), and a classical MLP ($2048$ hidden units). VQC-MLPNet achieves near-perfect accuracy within only a few epochs while maintaining the lowest loss, demonstrating its superior representation power and training efficiency.}
\label{fig:res_rep}
\end{figure}

\begin{table}[t]\footnotesize
\centering
\caption{Empirical results of quantum dot classification on the test dataset, showing the models' representation power with mean performance and uncertainties.}
\renewcommand{\arraystretch}{1.3}
\begin{tabular}{|c||c|c|c|c|}
\hline
\textbf{Models} 		& \textbf{Structures} 			& \textbf{Params.} 	& \textbf{Loss} 				& \textbf{Accuracy ($\%$)} \\
\hline
VQC-MLPNet 		& \small{20 Qubits, 6 depths} 	& $47.5$K 		& $0.0025 \pm 0.0002$ 		& $99.8 \pm 0.1$ \\
\hline
MLP 				& \small{$2048$ hidden units} 	& $5.13$M 		& $0.0146 \pm 0.0011$ 		& $99.5 \pm 0.2$ \\
\hline
TTN-VQC 		& \small{20 Qubits, 6 depths} 	& $3.28$K 		& $0.0173 \pm 0.0009$ 		& $99.6 \pm 0.2$ \\
\hline
VQC 			& \small{20 Qubits, 6 depths} 	& $402$			 & $0.4161 \pm 0.0023$ 		& $83.2 \pm 0.4$ \\
\hline
\end{tabular}
\label{tab:tab1}
\end{table}

Figure~\ref{fig:res_rep} and Table~\ref{tab:tab1} illustrate the representation capabilities of VQC-MLPNet compared to the classical MLP, standalone VQC, and TTN-VQC architectures evaluated on clean (noiseless) simulated quantum dot charge stability diagrams. The results indicate that VQC-MLPNet consistently achieves the lowest loss (0.00248) and highest accuracy ($99.8\% \pm 0.1\%$) across all epochs, rapidly converging to optimal performance. Although the classical MLP achieves a relatively high accuracy of $99.5\% \pm 0.2\%$, it exhibits higher training instability, likely due to its substantial parameter count (5.13 million), which is reflected in more noticeable fluctuations during training.

\vspace{0.5mm}

\hspace{6mm} TTN-VQC shows competitive initial performance, quickly reducing loss and increasing accuracy ($99.6\% \pm 0.2\%$), yet it converges notably slower and exhibits less training stability than VQC-MLPNet. The standalone VQC performs poorly overall, achieving the highest loss ($0.41605 \pm 0.0023$) and lowest accuracy ($83.2\% \pm 0.4\%$), consistent with theoretical predictions of limited representation power due to linearity and optimization difficulties.

\vspace{0.5mm}

\hspace{6mm} These empirical findings strongly support our theoretical analysis by demonstrating that integrating classical nonlinear activations with quantum-generated parameters significantly enhances representation capabilities and effectively reduces approximation error compared to both standalone VQC and TTN-VQC.

\vspace{3mm}

\item \textbf{Empirical Results on Generalization Power}:

\vspace{0.5mm}

Figure~\ref{fig:res_gen} and Table~\ref{tab:tab2} summarize the generalization performance of the evaluated models, tested using noisy quantum dot charge stability diagrams that simulate realistic experimental conditions. The VQC-MLPNet architecture achieves significantly lower test loss ($0.0596 \pm 0.0050$) and higher accuracy ($98.75\% \pm 0.40\%$) compared to other models, demonstrating its robustness and rapid convergence in realistic noisy scenarios.

\vspace{0.5mm}

\hspace{6mm} In contrast, the classical MLP model shows slower convergence, higher test loss ($0.2976 \pm 0.0100$), and substantially lower accuracy ($91.50\% \pm 1.00\%$), highlighting its sensitivity to noise-induced data fluctuations. TTN-VQC initially improves quickly but plateaus at a relatively high loss ($0.2873 \pm 0.0080$) and lower accuracy ($90.75\% \pm 1.00\%$), emphasizing optimization difficulties in fully quantum-based structures. The standalone VQC exhibits severe limitations, reflected by its consistently high loss ($0.6035 \pm 0.0040$) and significantly reduced accuracy ($72.25\% \pm 2.00\%$), underscoring its limited generalization capabilities.

\vspace{0.5mm}

\hspace{6mm} The presented empirical outcomes strongly validate our theoretical predictions, clearly demonstrating that VQC-MLPNet substantially improves generalization power, robustness, and training efficiency compared to purely classical neural networks and existing hybrid quantum-classical architectures under realistic noisy environments.

\begin{figure}
\centerline{\epsfig{figure=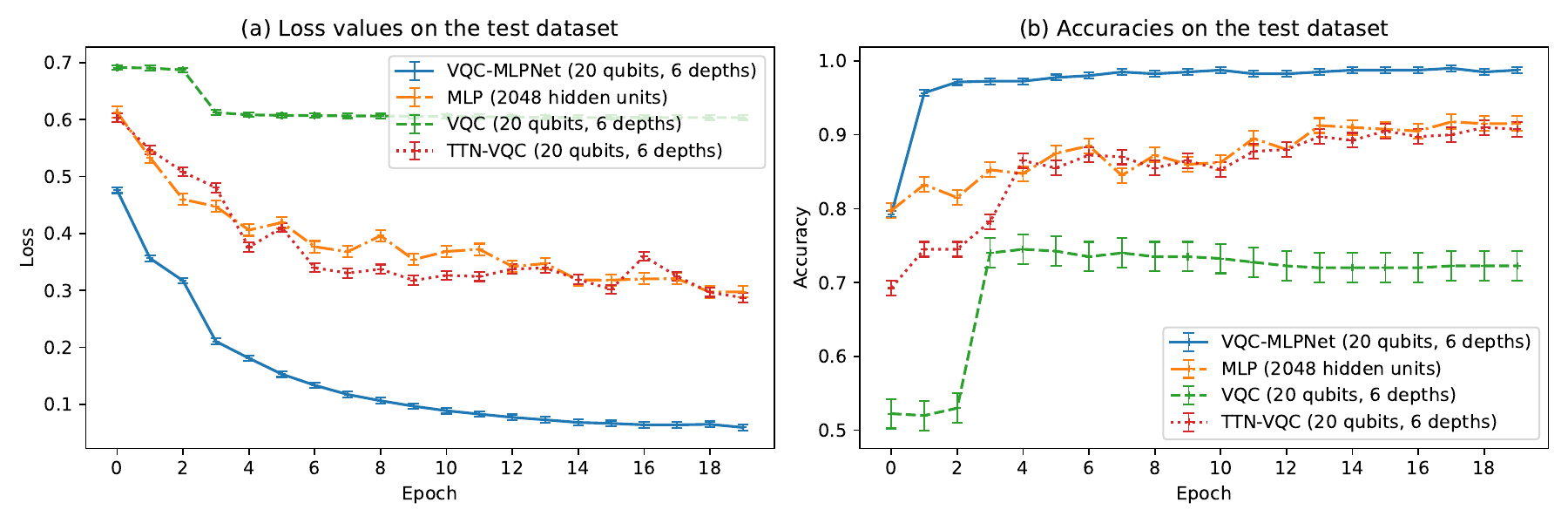, width=160mm}}
\caption{\textbf{Empirical results of quantum dot classification on the test dataset to evaluate models’ generalization power}. (a) Test loss and (b) test accuracy over 20 epochs for VQC-MLPNet ($20$ qubits, $6$ layers), TTN-VQC ($20$ qubits, $6$ layers), a plain VQC ($20$ qubits, $6$ layers), and a classical MLP ($2048$ hidden units). VQC-MLPNet achieves the lowest losses and highest accuracies, demonstrating stronger generalization power than the MLP, TTN-VQC, and plain VQC. } 
\label{fig:res_gen}
\end{figure}

\begin{table}[t]\footnotesize
\centering
\caption{Empirical results of quantum dot classification on the test dataset, showing the models' generalization power with mean performance and uncertainties.}
\renewcommand{\arraystretch}{1.3}
\begin{tabular}{|l||c|c|c|c|}
\hline
\textbf{Models} & \textbf{Structures} & \textbf{Params.} & \textbf{Loss} & \textbf{Accuracy (\%)} \\
\hline
VQC\textendash MLPNet & \small 20 qubits, 6 depths & $47.46\,\mathrm{K}$ & $0.0596 \pm 0.0050$ & $98.75 \pm 0.40$ \\
\hline
MLP                   & \small 2048 hidden units   & $5.13\,\mathrm{M}$ & $0.2976 \pm 0.0100$ & $91.50 \pm 1.00$ \\
\hline
TTN\textendash VQC    & \small 20 qubits, 6 depths & $3{,}283$          & $0.2873 \pm 0.0080$ & $90.75 \pm 1.00$ \\
\hline
VQC                   & \small 20 qubits, 6 depths & $402$              & $0.6035 \pm 0.0040$ & $72.25 \pm 2.00$ \\
\hline
\end{tabular}
\label{tab:tab2}
\end{table}

\vspace{3mm}

\item \textbf{The Effect of Qubit Number on Representation Power}: 

\vspace{0.5mm}

Figure~\ref{fig:res_rep_qubits} illustrates the empirical results assessing how varying the number of qubits influences the representation power of the VQC-MLPNet model, related to our theoretical analysis of the approximation error. Models with different numbers of qubits ($6$, $12$, and $20$ qubits), each with a fixed circuit depth of $6$, are compared using clean, noiseless quantum dot charge stability diagrams.

\vspace{0.5mm}

\hspace{6mm} From the results in Figure~\ref{fig:res_rep_qubits}, we observe that increasing the number of qubits significantly enhances the model's representation capability. Specifically, the $20$-qubit VQC-MLPNet achieves the lowest test loss and the highest accuracy, rapidly converging within the initial epochs and maintaining consistently high performance throughout the training process. The $12$-qubit model demonstrates competitive performance, yet it shows slightly higher loss and lower accuracy than the $20$-qubit model, reflecting a reduction in quantum expressivity. The $6$-qubit model displays the slowest convergence, highest loss, and more pronounced fluctuations in accuracy, clearly indicating limited representation power due to insufficient quantum resources. 

\begin{figure}
\centerline{\epsfig{figure=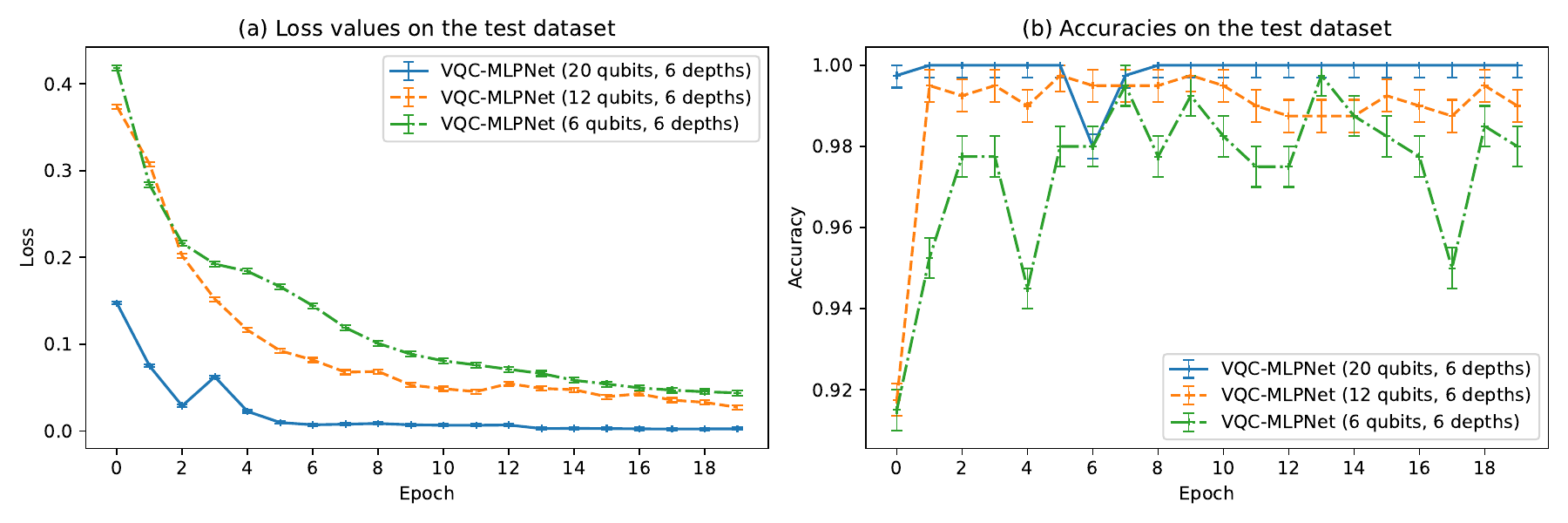, width=160mm}}
\caption{\textbf{Empirical results of quantum dot classification on the test dataset using VQC-MLPNet with fixed depth (6) but different qubit counts (6, 12, and 20), connected to the theoretical approximation error}. (a) Test loss curves show that larger qubit numbers accelerate convergence and reduce the final loss. (b) Accuracy curves demonstrate that more qubits yield stronger generalization. These results highlight that expanding the quantum feature space via additional qubits enhances the representation power and generalization ability of VQC-MLPNet.}
\label{fig:res_rep_qubits}
\end{figure}

\vspace{0.5mm}

\hspace{6mm} These experimental observations align closely with our theoretical approximation error analysis, confirming that the approximation power of VQC-MLPNet improves exponentially as the number of qubits increases. This empirically supports our theoretical predictions regarding the critical role of qubit numbers in enhancing the representation capabilities of quantum-enhanced models.

\vspace{3mm}

\item \textbf{The Effect of Quantum Circuit Depth on Representation Power}: 

\vspace{0.5mm}

Figure~\ref{fig:res_rep_depths} presents empirical results examining how varying the circuit depth of VQC-MLPNet affects its representation power, corresponding to our theoretical approximation error analysis. We compare VQC-MLPNet models with circuit depths of $1$, $3$, and $6$, each utilizing a fixed number of $20$ qubits, trained and tested on clean, noiseless quantum dot charge stability diagrams. 

\begin{figure}
\centerline{\epsfig{figure=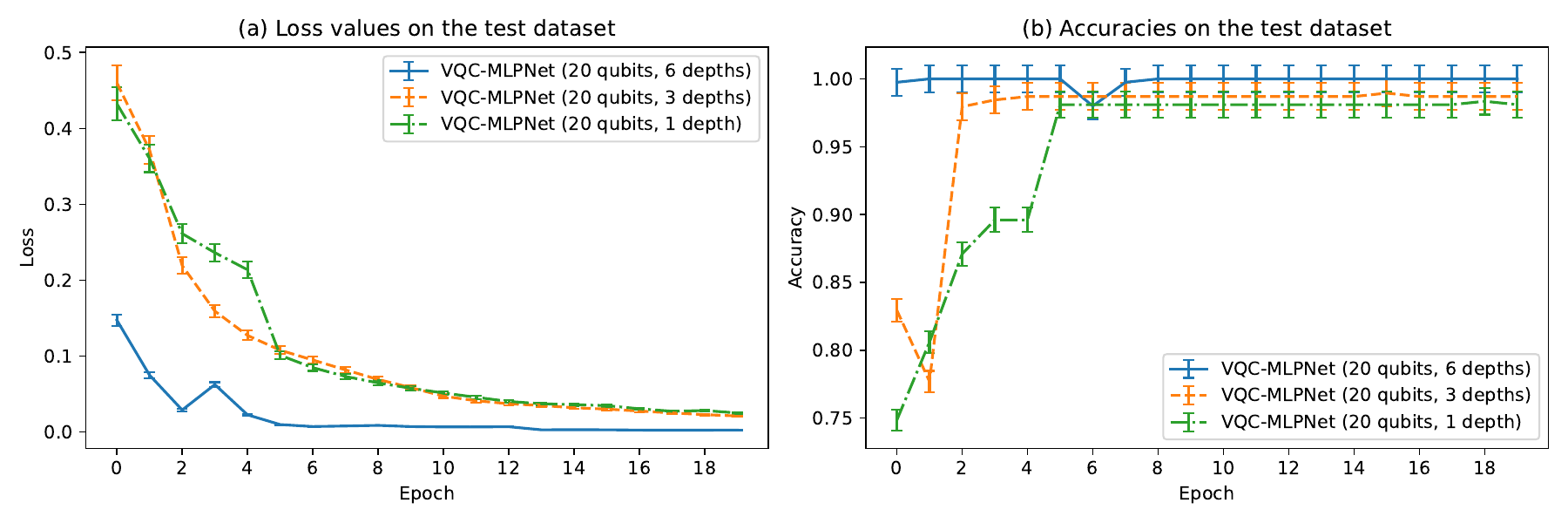, width=160mm}}
\caption{\textbf{Empirical results of quantum dot classification on the test dataset to demonstrate the impact of varying quantum circuit depths on the representation power of VQC-MLPNet, related to theoretical approximation error analysis}. Models with different circuit depths ($1$, $3$, and $6$) are evaluated using clean quantum dot charge stability diagrams. (a) Test loss and (b) accuracy curves indicate that deeper circuits significantly improve convergence speed, achieve lower approximation error, and yield higher accuracy, thus enhancing the overall representation capability of the model.}
\label{fig:res_rep_depths}
\end{figure}

\vspace{0.5mm}

\hspace{6mm} As depicted, deeper quantum circuits significantly enhance the representation capacity of the VQC-MLPNet model. The model with the circuit depth $6$ achieves the lowest test loss and highest accuracy, quickly converging to nearly perfect performance. In contrast, shallower circuit models (depths 1 and 3) exhibit comparatively higher losses and lower accuracies, converging more slowly and with greater fluctuations, particularly at depth 1. These results demonstrate that increased quantum circuit depth substantially improves the expressivity and representation capability of the model.

\vspace{0.5mm}

\hspace{6mm} These findings empirically validate our theoretical analysis: deeper quantum circuits exponentially reduce approximation error, enabling superior model expressivity and more accurate representation of complex quantum data.

\vspace{3mm}

\item \textbf{Optimization Performance Analysis of VQC-MLPNet}: 

\vspace{0.5mm}

Figure~\ref{fig:res_gen_v2} illustrates the empirical results comparing the optimization performance of the original VQC-MLPNet model against a newly proposed structure, VQC-MLPNet\_v2, in which two distinct VQCs independently generate weight matrices $\hat{\textbf{W}}^{(1)}$ and $\hat{\textbf{W}}^{(2)}$. Both models are configured with $20$ qubits and circuit depth $6$, trained and evaluated using noisy quantum dot stability diagrams to simulate realistic scenarios. 

\begin{figure}
\centerline{\epsfig{figure=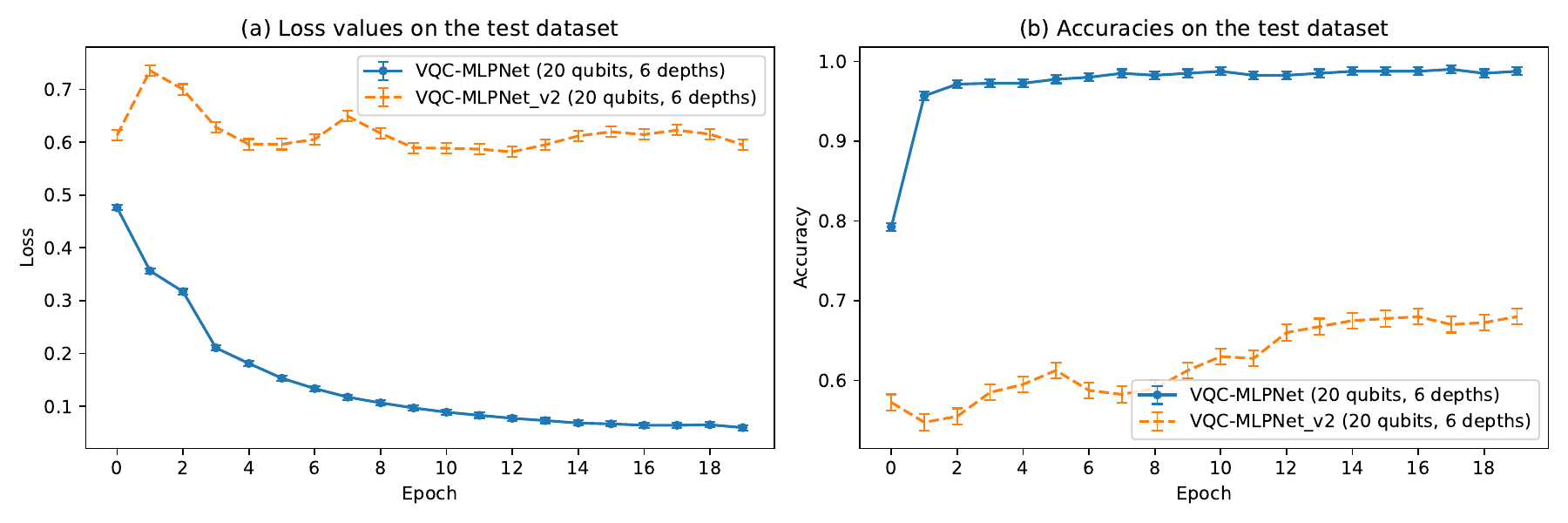, width=160mm}}
\caption{\textbf{Empirical results of quantum dot classification on the test dataset to compare optimization performance between the original VQC-MLPNet and the newly proposed VQC-MLPNet\_v2 architecture, where two independent VQCs generate the MLP's weight matrices $\hat{\textbf{W}}^{(1)}$ and $\hat{\textbf{W}}^{(2)}$}. Both models are configured with $20$ qubits and a circuit depth of $6$, trained on noisy quantum dot charge stability diagrams. (a) Test loss and (b) accuracy plots demonstrate that the original VQC-MLPNet significantly outperforms VQC-MLPNet$\_$v2, highlighting that the hybrid NTK kernel structure of VQC-MLPNet facilitates better optimization efficiency and stable convergence.}
\label{fig:res_gen_v2}
\end{figure}

\vspace{0.5mm}

\hspace{6mm} As Figure~\ref{fig:res_gen_v2}(a) shows, VQC-MLPNet demonstrates substantially lower test loss, rapidly converging to near-optimal values within a few epochs. In contrast, VQC-MLPNet\_v2 exhibits noticeably higher loss and unstable convergence behavior, indicating difficulties in effective optimization. Figure~\ref{fig:res_gen_v2}(b) further confirms this finding, with VQC-MLPNet quickly achieving and consistently maintaining significantly higher classification accuracy, whereas VQC-MLPNet\_v2 struggles to improve and shows limited accuracy throughout training.

\vspace{0.5mm}

\hspace{6mm} These results strongly align with our theoretical optimization analysis based on the NTK. More specifically, the superior performance of VQC-MLPNet empirically confirms that hybrid architectures, which leverage a balanced combination of quantum and classical kernels, yield a better-conditioned NTK, facilitating efficient optimization and stable convergence. Conversely, the suboptimal performance of VQC-MLPNet\_v2 suggests that independently optimized quantum circuits may lead to a less favorable NTK structure, resulting in slower convergence and instability in practical training scenarios.

\vspace{4mm}

\item \textbf{Empirical Results Under Realistic IBM Quantum Noises}

\vspace{0.5mm}

To further validate the robustness and applicability of the proposed VQC-MLPNet framework in practical quantum computing environments, we simulated realistic noise effects based on the IBM quantum hardware model~\cite{cross2019validating, resch2021benchmarking}. We introduced amplitude damping (ADR) and phase damping (PDR)~\cite{temme2017error} noises at different levels into the quantum circuits of our VQC-MLPNet model during training. The empirical results illustrated in Figure~\ref{fig:ibm} demonstrate that VQC-MLPNet maintains stable and accurate performance despite quantum noise. Even at relatively high noise levels (ADR = $0.01$ and PDR = $0.01$), the loss curve rapidly converged to low values, and accuracy remained comparable to that in the noise-free scenario, highlighting the architecture's inherent resilience to quantum hardware imperfections.

\vspace{0.5mm}

\hspace{6mm} This notable robustness arises primarily from the hybrid quantum-classical structure of VQC-MLPNet, where the classical MLP effectively compensates and mitigates quantum noise-induced inaccuracies, preserving overall model stability and predictive power. These findings underscore the practical viability of VQC-MLPNet for quantum machine learning tasks in NISQ devices, reinforcing its potential as a key approach toward achieving quantum advantages in realistic scenarios.

\begin{figure}
\centerline{\epsfig{figure=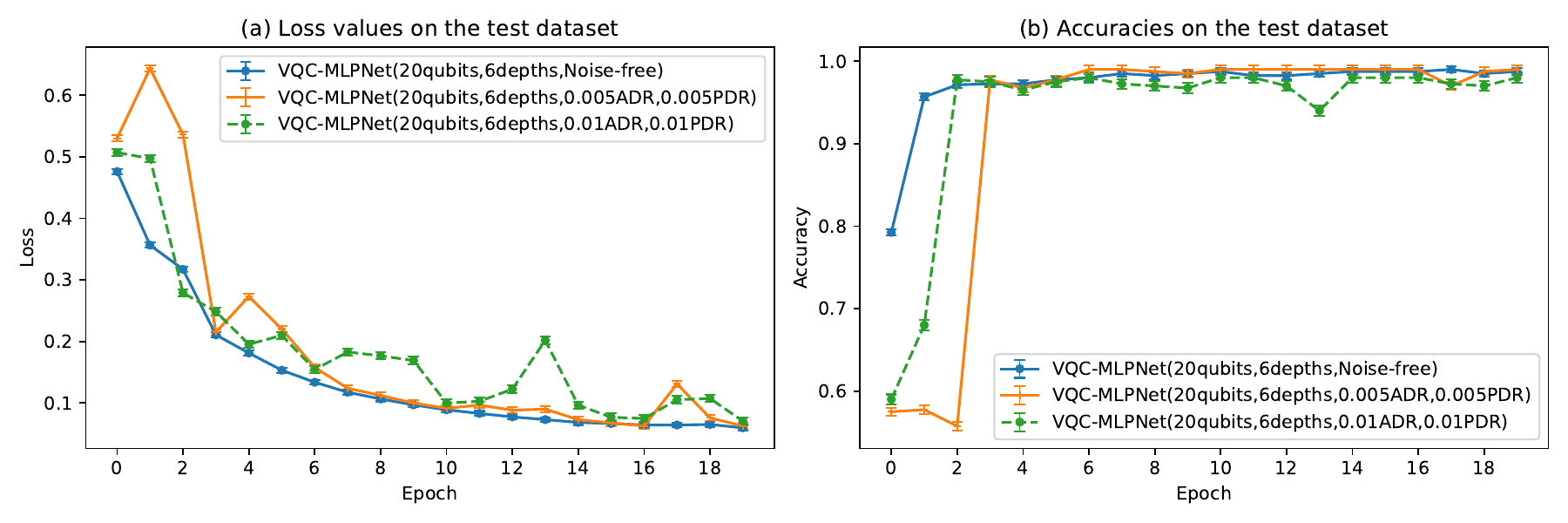, width=160mm}}
\caption{\textbf{Empirical results of quantum dot classification on the test dataset for VQC-MLPNet under different quantum noise levels}. (a) Test loss and (b) test accuracy over 20 epochs show a clear noise–performance trade-off: the noise-free model converges fastest with the lowest final loss and near-perfect accuracy, mild noise slows convergence with a small accuracy drop, and heavier noise yields the highest loss and lowest accuracy.}
\label{fig:ibm}
\end{figure}

\subsection{Empirical Results of Genome TFBS Prediction}

Next, we evaluate the proposed VQC-MLPNet's performance through a genome transcription factor binding sites (TFBS) prediction task. TFBS prediction is a binary classification problem determining whether specific DNA sequences contain particular binding-site patterns. We specifically focus on transcription factor JunD (TF JunD, InterPro entry IPR029823), a crucial activator of the protein-1 family involved in regulating gene transcription across various biological processes.

\vspace{0.5mm}

\hspace{6mm} Our dataset comprises $1,900$ genome-wide DNA segments collected from national experiments that identify TF JunD binding sites across all $22$ human chromosomes. We partition the dataset into $1,600$ segments for training and $300$ for testing. Each DNA sequence segment consists of $101$ bases labeled according to the presence or absence of the JunD binding site. Each base comprises four nucleotides (A, C, T, or G), so every segment is represented by a $404$-dimensional feature vector.

\vspace{0.5mm}

\hspace{6mm} We compare the generalization capabilities of the proposed VQC-MLPNet model against standalone VQC, classical MLP, and hybrid TTN-VQC models. Each quantum-based model is executed with $20$ qubits, a circuit depth of $6$, and measurement statistics collected from 4,096 shots. The classical MLP baseline employs a hidden-layer width of 2,048 units, yielding approximately 5.13 million parameters, whereas our proposed VQC-MLPNet reduces the parameter count to approximately 47.46 thousand. All models are trained for 30 epochs using the cross-entropy loss function optimized with Adam at a fixed learning rate of 0.001. To ensure fair comparisons, we adopt identical random seeds ($5$ in total) and parameter initializations drawn from a normal distribution. Quantum simulations are carried out on two RTX 4090 GPUs with a combined memory capacity of 48 GB.

\vspace{0.5mm}

\hspace{6mm} Figure \ref{fig:res_tf_res} and Table~\ref{tab:tab3} illustrate the empirical results of our TFBS prediction experiments. Figure~\ref{fig:res_tf_res}(a) demonstrates that the VQC-MLPNet achieves significantly lower test loss than classical MLP, TTN-VQC, and standalone VQC, rapidly converging to optimal values. The superior generalization capability of VQC-MLPNet is also evident in Figure~\ref{fig:res_tf_res}(b), where it consistently exhibits higher accuracy ($93.3\% \pm 0.06\%$) and more stable convergence across epochs compared to classical MLP ($90.5\% \pm 0.04\%$), TTN-VQC ($90.7\% \pm 0.05\%$), and standalone VQC ($68.7\% \pm 0.06\%$). Notably, the standalone VQC struggles with limited accuracy improvement due to its restricted representation capability, inherent in its linear structure, and optimization challenges. While the classical MLP and TTN-VQC perform better than the standalone VQC, they still exhibit noticeable fluctuations and lower accuracy compared to VQC-MLPNet.

\begin{figure}
\centerline{\epsfig{figure=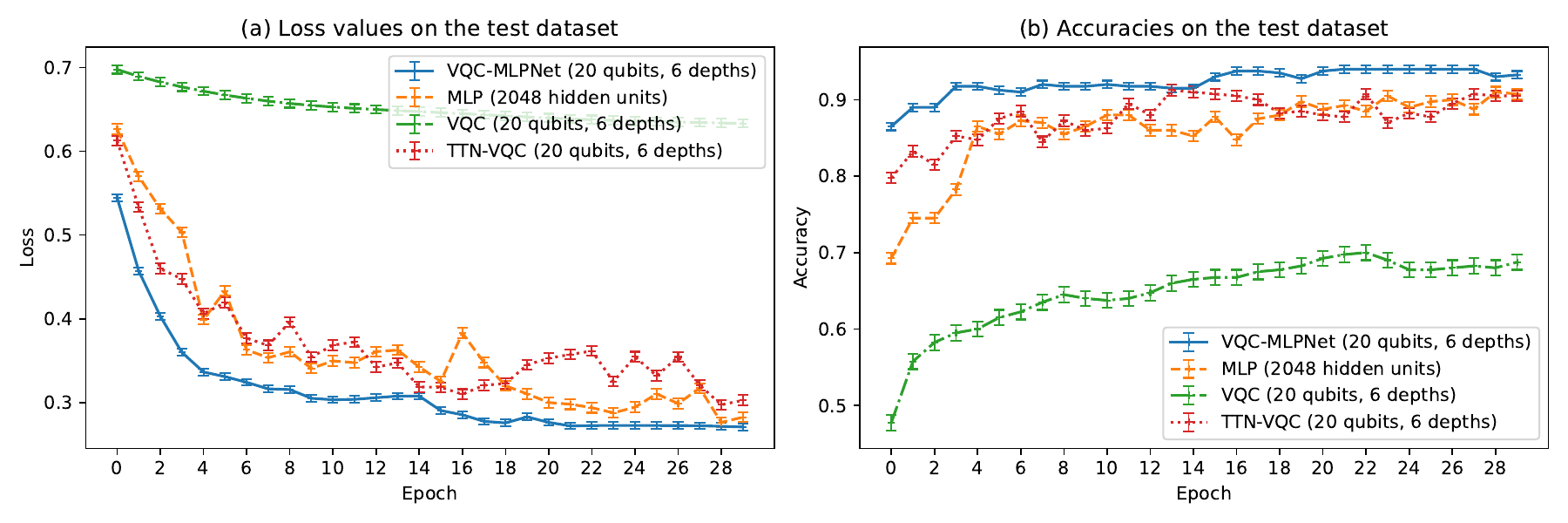, width=160mm}}
\caption{\textbf{Empirical results of genome TFBS prediction on the test dataset to compare the generalization performance of VQC-MLPNet, classical MLP, standalone VQC, and TTN-VQC}. (a) Loss and (b) accuracy over $30$ epochs comparing four models: VQC-MLPNet (20 qubits, $6$ depths), a classical MLP (2048 hidden units), a plain VQC (20 qubits, $6$ depths), and TTN-VQC (20 qubits, $6$ depths). VQC-MLPNet attains the highest test accuracy throughout training with a competitive final loss, while the plain VQC lags in both metrics; TTN-VQC and the MLP perform in between. These trends highlight the generalization advantage of the hybrid VQC-MLPNet architecture for TFBS classification.}
\label{fig:res_tf_res}
\end{figure}

\begin{table}[tpbh]\footnotesize
\center
\caption{Empirical results of genome TFBS prediction on the test dataset, showing the models' generalization power with mean performance and uncertainties.}
\renewcommand{\arraystretch}{1.3}
\begin{tabular}{|c||c|c|c|c|}
\hline
\textbf{Models} & \textbf{Structures} & \textbf{Params.} & \textbf{Loss} & \textbf{Accuracy ($\%$)} \\
\hline
VQC-MLPNet & \small{20 Qubits, 6 depths} & $47.46$K & $0.2708 \pm 0.0040$ & $93.3 \pm 0.06$ \\
\hline
MLP & \small{$2048$ hidden units} & $5.13$M & $0.2825 \pm 0.0060$ & $90.5 \pm 0.04$ \\
\hline
TTN\textendash VQC & \small{20 Qubits, 6 depths} & $3283$ & $0.3026 \pm 0.0060$ & $90.7 \pm 0.05$ \\
\hline
VQC & \small{20 Qubits, 6 depths} & $402$ & $0.6332 \pm 0.0100$ & $68.7 \pm 0.06$ \\
\hline
\end{tabular}
\label{tab:tab3}
\end{table}

\vspace{0.5mm}

\hspace{6mm} Figure~\ref{fig:res_tf_noise} further highlights the robustness of VQC-MLPNet under realistic quantum noise conditions simulated using IBM quantum noise with varying ADRs and PDRs. VQC-MLPNet experiences minimal performance degradation compared to the noise-free scenario, maintaining high predictive performance (above $90\%$) even under increased quantum noise levels. These results validate the inherent resilience and practical applicability of VQC-MLPNet in realistic quantum-enhanced genomic machine-learning scenarios within the NISQ era.

\begin{figure}
\centerline{\epsfig{figure=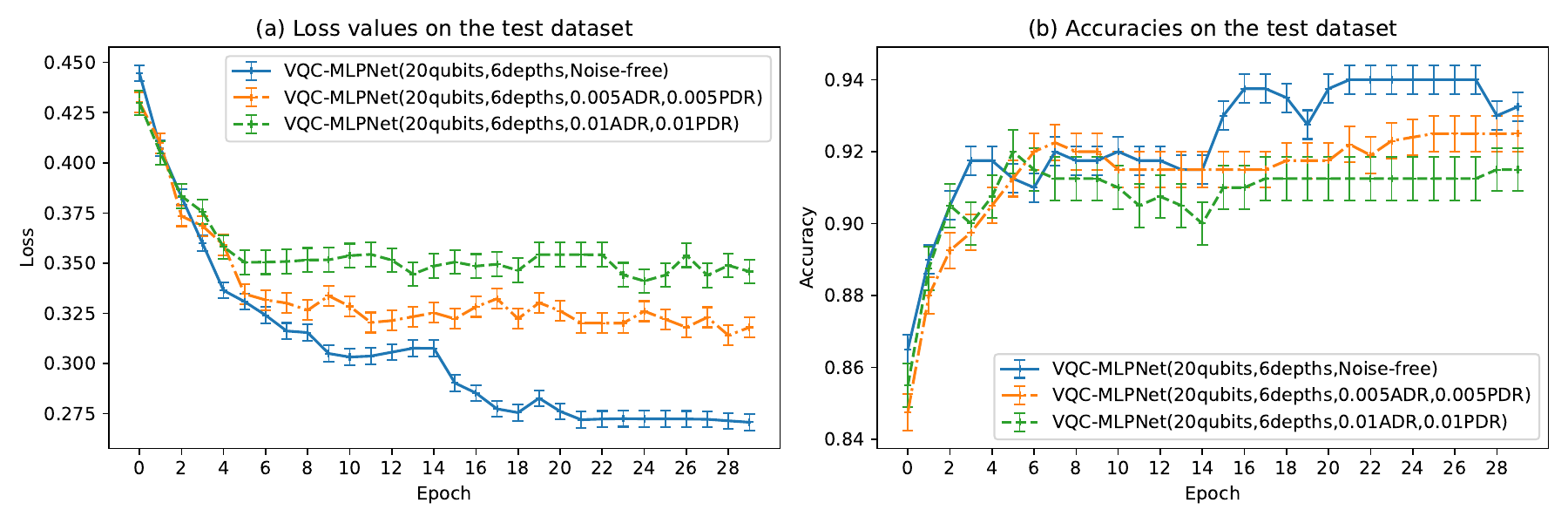, width=160mm}}
\caption{\textbf{Empirical results of genome TFBS prediction on the test dataset under IBM quantum noises}. (a) Test loss curves and (b) test accuracies over $30$ epochs for VQC-MLPNet with $20$ qubits and circuit depth $6$, comparing noise-free simulation with noisy simulations under ADR and PDR of 0.005 and 0.01, respectively.}
\label{fig:res_tf_noise}
\end{figure}

\vspace{0.5mm}

\hspace{6mm} In sum, our empirical findings strongly support our theoretical claims. They demonstrate the enhanced representation power, improved generalization ability, and robustness of VQC-MLPNet, reinforcing its potential as a robust, scalable hybrid quantum-classical architecture for practical genome analysis tasks.

\end{itemize}

\section{Discussion}

In this work, we introduce the VQC-MLPNet architecture, a hybrid quantum-classical neural network framework integrating classical MLPs with VQCs, to address significant limitations in quantum machine learning. Our theoretical analysis rigorously demonstrates that the VQC-MLPNet model systematically enhances approximation capabilities, generalization performance, and optimization stability relative to standalone VQCs and existing hybrid quantum-classical models, such as TTN-VQC. By decomposing the total learning error into approximation, uniform deviation, and optimization components, we derive explicit theoretical upper bounds illustrating the benefits of combining quantum expressivity with classical nonlinear architectures.

Our empirical evaluations substantiate these theoretical predictions, demonstrating that VQC-MLPNet consistently outperforms competing approaches across diverse, challenging tasks, including semiconductor quantum dot classification and genome transcription factor binding site prediction. The experimental results demonstrate the enhanced training efficiency, superior expressivity, and improved generalization capability of the VQC-MLPNet model, particularly under realistic quantum noise conditions representative of near-term quantum computing scenarios.

The success of the VQC-MLPNet approach underscores the importance of effectively integrating classical neural network structures within quantum-enhanced parametric frameworks. This hybrid design paradigm addresses critical barriers in the practical deployment of quantum machine learning algorithms, including optimization instability and limited quantum circuit expressivity. VQC-MLPNet thus represents a significant step toward scalable, reliable, and practical quantum-assisted learning methods, leveraging classical nonlinear activation functions and efficient gradient-based optimization.

We further discuss three critical aspects of our proposed architecture: the rationale behind choosing the classical MLP as the classical component, the robustness of our model under realistic quantum noise scenarios, and the relevance of VQC-MLPNet to parameter-efficient learning in deep neural networks.

\subsection{The Choice of MLP for The Hybrid Quantum-Classical Model}

The decision to adopt an MLP as the classical backbone of our hybrid architecture is driven by key practical and theoretical considerations. For one thing, an MLP equipped with a sufficiently large number of neurons can universally approximate complex functions represented by deep neural networks or convolutional neural networks. Thus, the VQC-MLPNet architecture inherits the flexibility and representation power of more sophisticated neural network structures while maintaining a straightforward and computationally efficient form. Furthermore, integrating classical MLPs with quantum circuits significantly enhances the expressivity, trainability, and generalization capabilities of VQCs. These advantages establish the classical MLP as a highly effective choice for boosting VQC performance.

\subsection{Practicality and Robustness of VQC-MLPNet Under Realistic Quantum Noise}

In evaluating the practical applicability of our proposed VQC-MLPNet architecture, we conducted simulations incorporating realistic IBM quantum noise models~\cite{resch2021benchmarking}. Specifically, we applied amplitude and phase damping channels characterized by their respective damping rates (ADR and PDR)~\cite{temme2017error}. These quantum noise simulations accurately reflect decoherence mechanisms frequently encountered in real-world quantum computing devices such as superconducting qubit systems. Therefore, our noise modeling provides a reliable assessment of the practical robustness of VQC-MLPNet.

Our experimental results confirm that the VQC-MLPNet model is remarkably robust under these realistic noise conditions. Even at moderate damping levels (e.g., ADR and PDR up to $0.01$), VQC-MLPNet maintains excellent classification accuracy and effectively converges during training. This resilience stems from the hybrid architecture, where the classical MLP ensures stable nonlinear processing, while the quantum circuit enhances the model's expressivity without overexposing it to detrimental quantum noise. Furthermore, the layered structure of VQC-MLPNet partially decouples the classical learning process from quantum noise effects, mitigating the optimization difficulties frequently observed in purely quantum circuits. This robustness significantly enhances the model's practical utility in contemporary and near-term quantum computing platforms.

\subsection{Connection to The Parameter-Efficient Learning of Deep Neural Networks}

The VQC-MLPNet architecture significantly contributes to parameter-efficient learning in neural networks~\cite{liu2024quantum}. One notable strength of the VQC-MLPNet model is its ability to retain high expressivity and strong generalization power while drastically reducing the required parameters. Traditional deep neural networks with large hidden layers often suffer from inefficiencies due to the exponentially increasing number of parameters needed to model complex data relationships. Such over-parameterization can lead to training instability, overfitting, and poor generalization, particularly in scenarios with limited data.

In contrast, VQC-MLPNet addresses these issues by incorporating quantum circuits as compact, expressive parameter generators within a classical MLP framework. Quantum circuits inherently offer substantial representational efficiency, enabling the encoding of complex functional relationships in a more compact manner. Consequently, VQC-MLPNet achieves significant reductions in model parameters (e.g., approximately 100× fewer parameters than classical counterparts) without compromising predictive performance. This hybrid approach offers a powerful and general strategy for building efficient deep learning models, achieving the dual objectives of reduced complexity and enhanced generalization capability.

\subsection{Comparative Insights into Other Emerging Quantum Architectures}

Recent developments in quantum machine learning have introduced several innovative quantum-inspired architectures, such as the Quantum Convolutional Neural Networks (QCNNs)~\cite{cong2019quantum} and the Quantum Transformers (QTransformers)~\cite{kerenidis2024quantum, di2022dawn}. QCNNs utilize quantum convolutional and pooling layers to systematically extract hierarchical features directly within quantum circuits, offering compact parameterizations that are beneficial for structured quantum data, such as quantum states and many-body systems. In contrast, Quantum Transformers utilize attention mechanisms to harness quantum-inspired correlations, thereby effectively capturing long-range dependencies within quantum-enhanced feature spaces. While these models are emerging, they remain primarily quantum-centric, often facing challenges in optimization stability, noise resilience, and generalization on near-term quantum hardware. Our proposed VQC-MLPNet distinctively addresses these limitations by seamlessly integrating quantum circuit expressivity into a classical neural structure, thus improving nonlinear representation capability, optimization stability (as evidenced by favorable NTK properties), and robustness under realistic noise conditions. Consequently, VQC-MLPNet complements existing quantum-centric approaches by providing a practical and scalable hybrid solution suited explicitly for current quantum computing landscapes.

\subsection{Potential Broader Impacts of VQC-MLPNet}

Beyond the quantum dot classification and genomic prediction tasks demonstrated in this study, the proposed VQC-MLPNet framework holds significant promise for broader scientific domains that require complex, noise-resilient modeling. For instance, the approach could be naturally adapted to computational drug discovery, facilitating quantum-enhanced screening of large molecular datasets. The hybrid architecture's scalability and robustness against quantum noise in quantum chemistry could also enable more efficient molecular dynamics modeling or predicting electronic structures for complex materials. Furthermore, fields such as climate modeling, where data-driven predictions are challenged by inherent noise and high dimensionality, might benefit from our method's integration of quantum expressivity and classical nonlinearity. 




\section{Methods}

This section provides methodological support for the theoretical analyses presented in the main text. It introduces the NTK theory,  TTN, the TTN-VQC hybrid model, the cross-entropy loss function, and the proof sketches of our derived theoretical results, including the NTK-based analysis of VQC-MLPNet’s trainability. More detailed proofs are presented in the Appendix. 

\subsection{Introduction to Neural Tangent Kernel Theory}

The Neural Tangent Kernel (NTK) is a powerful theoretical tool that explains why deep neural networks can be trained efficiently despite their highly complex structure. At its core, NTK theory examines how small changes in a neural network’s parameters affect its predictions. If slight parameter adjustments consistently lead to stable and predictable changes in output, the network is considered “well-conditioned,” making training smoother and convergence faster. 

Formally, the NTK captures this stability by forming a kernel, a similarity measure, that describes the network’s behavior throughout training. A network with a “well-conditioned” NTK, whose smallest eigenvalue is not too small, will have stable gradients during training, enabling efficient optimization via gradient-based algorithms. Conversely, networks with poorly conditioned NTKs can exhibit instability and slow convergence, commonly referred to as optimization difficulties.

In our work, integrating quantum circuits with classical neural layers (VQC-MLPNet) improves the condition of the NTK, facilitating more efficient training and robust convergence. In particular, while our theoretical analysis builds upon the NTK framework, which formally assumes the infinite-width limit of the MLP component, our practical implementation uses a finite yet sufficiently large hidden width $M$. Specifically, we set $M = 2048$ in our experiments, a value large enough to closely approximate the NTK regime. Empirical results demonstrate that the model exhibits stable convergence and strong generalization behavior, which is consistent with the theoretical bounds derived under infinite-width assumptions. Thus, NTK theory helps us understand and characterize our model’s training dynamics and provides rigorous guarantees about its performance and generalization capabilities.

\subsection{TTN and TTN-VQC}

The tensor-train network (TTN)~\cite{oseledets2011tensor}, also known as the matrix product state (MPS)~\cite{perez2006matrix}, is a widely adopted tensor network architecture in machine learning and quantum physics. TTN efficiently represents high-dimensional tensors by factorizing them into a sequence of lower-dimensional tensors, usually matrices or tensors with three indices, arranged in a linear, chain-like structure. When combined with a VQC, TTN forms a hybrid quantum-classical architecture called TTN-VQC. The TTN effectively reduces input dimensionality in this hybrid setup, while the VQC enhances expressivity by introducing quantum-induced nonlinearity. This combination positions TTN-VQC as a robust framework for handling complex, high-dimensional data in quantum machine learning tasks. However, TTN-VQC addresses the common challenges and limitations of hybrid quantum-classical designs, including significant quantum resource requirements, linear representational constraints inherent to quantum circuits, and optimization difficulties due to complex parameter landscapes. Consequently, demonstrating VQC-MLPNet’s superiority over TTN-VQC directly underscores our proposed method's broader advancements and general applicability within the hybrid quantum-classical modeling paradigm.

\subsection{Cross-Entropy Loss Function}

The cross-entropy loss function~\cite{mao2023cross} is commonly used in classification tasks, quantifying the discrepancy between predicted probability distributions and the ground-truth labels. It is particularly effective for optimizing probabilistic models, including neural networks and quantum machine learning frameworks. In this study, we utilize the cross-entropy loss with gradient-based optimization methods, guiding model parameters toward improved classification performance by penalizing inaccurate predictions. We consistently apply this loss function to VQC-MLPNet, TTN-VQC, standalone VQC, and classical MLP models, ensuring a fair and unified assessment of their theoretical convergence and empirical performance.

\subsection{Proof Sketch of Theorem~\ref{thm:thm1}}

Consider a ground truth $\boldsymbol{y}$, a target operator $h^{*}$, an optimal MLP operator $f^{*}_{\rm mlp}$, and an VQC-MLPNet operator $f_{\boldsymbol{\theta}}^{*}$. The approximation error $\epsilon_{\rm app}$ can be upper bounded using the triangle inequality as: 
\begin{equation}
\begin{split}
\epsilon_{\rm app} &= \mathcal{R}(f_{\boldsymbol{\theta}^{*}}) - \mathcal{R}(h^{*}) \le \vert \mathcal{R}(f_{\boldsymbol{\theta}^{*}}) - \mathcal{R}(f_{\rm mlp}^{*}) \vert + \vert \mathcal{R}(f_{\rm mlp}^{*}) - \mathcal{R}(h^{*}) \vert. 
\end{split}
\end{equation}

Moreover, the expected risk $\mathcal{R}$ is taken as the cross-entropy loss function such that:
\begin{equation}
\begin{split}
\vert \mathcal{R}(f_{\boldsymbol{\theta}^{*}}) - \mathcal{R}(f_{\rm mlp}^{*}) \vert + \vert \mathcal{R}(f_{\rm mlp}^{*}) - \mathcal{R}(h^{*}) \vert \le L_{\rm ce} \lVert f_{\boldsymbol{\theta}^{*}} - f_{\rm mlp}^{*} \rVert_{2}  + L_{\rm ce} \lVert f_{\rm mlp}^{*} - h^{*} \rVert_{2}, 
\end{split}
\end{equation}
where $L_{\rm ce}$ is a Lipschitz constant induced by the gradient of the cross-entropy loss function with a softmax activation. 

As for the first term, we leverage Cybenco's universal approximation theory~\cite{cybenko1989approximation}, which leads to
\begin{equation}
\lVert f_{\boldsymbol{\theta}^{*}} - f_{\rm mlp}^{*} \rVert_{2} \le \mathcal{O}\left(\frac{1}{\sqrt{M}}\right). 
\end{equation}

Next, we define $f_{\rm mlp}^{*}$ and $f_{\boldsymbol{\theta}}^{*}$ operators explicitly:
\begin{equation}
f_{\rm mlp}^{*}(\textbf{x}) = \textbf{W}^{(2) \top} \sigma(\textbf{W}_{*}^{(1)\top} \textbf{x}), \hspace{4mm} f_{\boldsymbol{\theta}^{*}}(\textbf{x}) = \textbf{W}^{(2)\top} \sigma(\hat{\textbf{W}}^{(1)\top} \textbf{x}),
\end{equation}
and assuming a Lipschitz-continuous activation function $\sigma(\cdot)$, we apply the Cauchy-Swartz inequality, resulting in:
\begin{equation}
\Vert f_{\boldsymbol{\theta}^{*}} - f_{\rm mlp}^{*} \rVert_{2} \le \max\limits_{1\le j\le J} \left( \sum\limits_{m=1}^{M} \vert \textbf{w}^{(2)}_{m}(j)\vert \right) \lVert \textbf{x} \rVert_{2} \left\lVert \hat{\textbf{W}}^{(1)} - \textbf{W}_{*}^{(1)} \right\rVert_{2}.
\end{equation}

The key term $\left\lVert \hat{\textbf{W}}^{(1)} - \textbf{W}_{*}^{(1)} \right\rVert_{2}$ denotes an induced norm of matrix and can be upper bounded as the sum of encoding error $r_{\rm enc}$ and expressive error $r_{\rm exp}$:
\begin{equation}
\left\lVert \hat{\textbf{W}}^{(1)} - \textbf{W}_{*}^{(1)} \right\rVert_{2} \le  \underbrace{ \left\lVert \hat{\textbf{W}}^{(1)} - \hat{\textbf{W}}^{(1)}_{Q} \right\rVert_{2} }_{r_{\rm enc}} +  \underbrace{ \left\lVert \hat{\textbf{W}}^{(1)}_{Q} - \textbf{W}_{*}^{(1)} \right\rVert_{2}}_{r_{\rm exp}},
\end{equation}
where $\hat{\textbf{W}}^{(1)}_{Q}$ is defined as: 
\begin{equation}
\hat{\textbf{W}}^{(1)}_{Q} = \left[ \vert \hat{\textbf{w}}^{(1)}_{1} \rangle, \vert \hat{\textbf{w}}^{(1)}_{2} \rangle, ..., \vert \hat{\textbf{w}}^{(1)}_{D} \rangle \right]. 
\end{equation}

In particular, $\hat{\textbf{W}}_Q$ is composed of normalized state amplitudes, which are obtained via statistical sampling of measurement outcomes on a quantum device rather than being physically accessible, consistent with the Born rule. 

On the one hand, the amplitude encoding implies: 
\begin{equation}
r_{\rm enc} \le \mathcal{O}\left(	\frac{1}{2^{\beta U}} \right),
\end{equation}
with a scaling constant $\beta \in (0, \frac{1}{2})$ and the number of qubits $U$. On the other hand, since the target matrix $W^{*}$ is smooth, using the Fourier expansion and exponential decay of coefficients~\cite{schuld2021effect, stein2011fourier}, the expressive error satisfies:
\begin{equation}
r_{\rm exp} \le \mathcal{O}\left( e^{-\alpha L} \right), 
\end{equation}
with circuit depth $L$ and problem-dependent constant $\alpha > 0$. Combining these results yields the approximation error bound as Eq. (\ref{eq:app_error}).

\subsection{Proof Sketch of Theorem~\ref{thm:thm2}}

Based on the statistical learning theory, we utilize empirical Rademacher complexity~\cite{vershynin2018high, mohri2018foundations} to derive an upper bound for the uniform deviation error  $\epsilon_{\rm dev}$:
\begin{equation}
\epsilon_{\rm dev} \le 2 \sup\limits_{\boldsymbol{\theta} \in \Theta}\left\vert \hat{\mathcal{R}}(f_{\boldsymbol{\theta}}) - \mathcal{R}(f_{\boldsymbol{\theta}}) \right\vert \le 2\hat{\mathcal{C}}_{S}(\mathcal{F}_{\rm vm}), 
\end{equation}

The empirical Rademacher complexity $\hat{\mathcal{C}}_{S}(\mathcal{F}_{\rm vm})$ is defined as: 
\begin{equation}
\hat{\mathcal{C}}_{S}(\mathcal{F}_{\rm vm}) = \frac{1}{|S|} \mathbb{E}_{\boldsymbol{\sigma}}\left[ \sup\limits_{\lVert \textbf{w}_{m}^{(2)} \rVert_{1} \le \Lambda, \lVert \hat{\textbf{w}}^{(1)}_{m} \rVert_{2} \le \Lambda_{Q}} \sum\limits_{i=1}^{|S|} \sigma_{i} \sum\limits_{m=1}^{M} \textbf{w}_{m}^{(2)} \sigma(\hat{\textbf{w}}^{(1)}_{m} \cdot \textbf{x}_{i})	 \right]. 
\end{equation}

Employing techniques of expanding the supremum and bounding used in~\cite{qi2023theoretical}, we obtain: 
\begin{equation}
\hat{\mathcal{C}}_{S}(\mathcal{F}_{\rm vm}) \le \frac{\Lambda \Lambda_{Q}r}{\sqrt{\vert S\vert}},
\end{equation}
where $\Lambda_{Q}$ quantifies quantum circuit complexity and is influenced by circuit depth $L$. In particular, in line with insights from quantum circuit models and variational quantum algorithms studies~\cite{preskill2018quantum, cerezo2021variational, abbas2021power}, we assume that the quantum feature norm $\Lambda_{Q}$ scales as $\mathcal{O}(\sqrt{L})$, motivating a refined upper bound as:
\begin{equation}
\hat{\mathcal{C}}_{S}(\mathcal{F}_{\rm vm}) \le \frac{\Lambda \sqrt{L} r}{\sqrt{\vert S \vert}}. 
\end{equation}

\subsection{The NTK Technique for VQC-MLPNet's Trainability}

To analyze the trainability of the VQC-MLPNet model, we consider the operator $f_{\rm vm}$ with a quantum-enhanced weight $\hat{\textbf{W}}^{(1)} = f_{\rm lin} \circ f_{\rm vqc} \circ f_{\rm ae}(\textbf{W}^{(1)})$, where $f_{\rm vqc}$ is parameterized by $\boldsymbol{\theta}_{\rm vqc} = \{ \alpha_{1:U}, \beta_{1:U}, \gamma_{1:U} \}$, and the final layer uses classical weights $\boldsymbol{\theta}_{W^{(2)}}$. The VQC-MLPNet operator is given by:
\begin{equation}
f_{\boldsymbol{\theta}}(\textbf{x}) = \frac{1}{\sqrt{M}} \sum\limits_{m=1}^{M} \textbf{w}_{m}^{(2)} \sigma\left( \langle \hat{\textbf{w}}_{m}^{(1)}, \textbf{x} \rangle \right).
\end{equation}

\vspace{1.5mm}
We define the NTK for VQC-MLPNet, $\mathcal{K}_{\rm vm}$, as a linear combination of the VQC and the classical weight kernel: 
\begin{equation}
\begin{split}
\mathcal{K}_{\rm vm}&= \mathcal{K}_{\rm vqc} + \mathcal{K}_{W^{(2)}}.	 \\
\end{split}
\end{equation}

\vspace{1.5mm}

Since the VQC component of the NTK can be decomposed into contributions from the parameterized quantum gates, for any two data vectors $\textbf{x}_{1}$ and $\textbf{x}_{2}$, the related NTK $\mathcal{K}_{\rm vqc}(\textbf{x}_{1}, \textbf{x}_{2})$ can be further decomposed as: 
\begin{equation}
\mathcal{K}_{\rm vqc}(\textbf{x}_{1}, \textbf{x}_{2}) = \mathcal{K}_{\rm vqc}^{(\boldsymbol{\alpha})}(\textbf{x}_{1}, \textbf{x}_{2}) + \mathcal{K}_{\rm vqc}^{(\boldsymbol{\beta})}(\textbf{x}_{1}, \textbf{x}_{2}) +  \mathcal{K}_{\rm vqc}^{(\boldsymbol{\gamma})}(\textbf{x}_{1}, \textbf{x}_{2}).
\end{equation}

\vspace{1.5mm}

Each component is computed over the quantum-enhanced features. For instance, given a constant $C_{\boldsymbol{\alpha}}$, the $\boldsymbol{\alpha}$-parameter NTK contribution is:
\begin{equation}
\begin{split}
 \mathcal{K}_{\rm vqc}^{(\boldsymbol{\alpha})}(\textbf{x}_{1}, \textbf{x}_{2})
 &=  \frac{1}{M} \sum\limits_{m=1}^{M} \left(\textbf{w}_{m}^{(2)} \right)^{2} \sigma^{'}\left(\left\langle \hat{\textbf{w}}_{m}^{(1)}, \textbf{x}_{1} \right\rangle\right) \sigma^{'}\left(\left\langle \hat{\textbf{w}}_{m}^{(1)}, \textbf{x}_{2} \right\rangle\right) \left\langle \textbf{x}_{1}, C_{\boldsymbol{\alpha}} \textbf{x}_{2}\right\rangle,
\end{split}
\end{equation}
\noindent with similar expressions for $\mathcal{K}_{\rm vqc}^{(\boldsymbol{\beta})}$ and $\mathcal{K}_{\rm vqc}^{(\boldsymbol{\gamma})}$ using constants $C_{\boldsymbol{\beta}}$ and $C_{\boldsymbol{\gamma}}$ as follows:

\begin{equation}
\begin{split}
\label{eq:beta}
 \mathcal{K}_{\rm vqc}^{(\boldsymbol{\beta})}(\textbf{x}_{1}, \textbf{x}_{2}) 
 &=  \frac{1}{M} \sum\limits_{m=1}^{M} \left(\textbf{w}_{m}^{(2)} \right)^{2} \sigma^{'}\left(\left\langle \hat{\textbf{w}}_{m}^{(1)}, \textbf{x}_{1} \right\rangle\right) \sigma^{'}\left(\left\langle \hat{\textbf{w}}_{m}^{(1)}, \textbf{x}_{2} \right\rangle\right) \left\langle \textbf{x}_{1}, C_{\boldsymbol{\beta}} \textbf{x}_{2}\right\rangle,
\end{split}
\end{equation}

\begin{equation}
\label{eq:gamma}
\begin{split}
 \mathcal{K}_{\rm vqc}^{(\boldsymbol{\gamma})}(\textbf{x}_{1}, \textbf{x}_{2}) 
 &=  \frac{1}{M} \sum\limits_{m=1}^{M} \left(\textbf{w}_{m}^{(2)} \right)^{2} \sigma^{'}\left(\left\langle \hat{\textbf{w}}_{m}^{(1)}, \textbf{x}_{1} \right\rangle\right) \sigma^{'}\left(\left\langle \hat{\textbf{w}}_{m}^{(1)}, \textbf{x}_{2} \right\rangle\right) \left\langle \textbf{x}_{1}, C_{\boldsymbol{\gamma}} \textbf{x}_{2}\right\rangle.
\end{split}
\end{equation}

Similarly, the classical linear layer NTK is given by:
\begin{equation}
\begin{split}
\mathcal{K}_{W^{(2)}}(\textbf{x}_{1}, \textbf{x}_{2}) 
= \frac{1}{M} \sum\limits_{m=1}^{M} \sigma^{'}\left( \langle \hat{\textbf{w}}_{m}^{(1)}, \textbf{x}_{1} \rangle \right) \sigma^{'}\left( \langle \hat{\textbf{w}}_{m}^{(1)}, \textbf{x}_{2} \rangle \right). 
\end{split}
\end{equation}

In the infinite-width limit $M \rightarrow \infty$, the NTKs converge to their expected values under the initialization distribution. This yields constant kernels for each component (Eqs. (\ref{eq:t1}) and (\ref{eq:t2})), enabling analytical tractability of convergence behavior. 

\begin{equation}
\label{eq:t1}
\small{\mathcal{K}_{\rm vqc}^{(\boldsymbol{\alpha})}(\textbf{x}_{1}, \textbf{x}_{2})  \xrightarrow{M \rightarrow \infty} \mathbb{E}_{\left(\boldsymbol{\theta}_{\rm vqc}, \boldsymbol{\theta}_{W^{(2)}}\right)} \left[ (\textbf{w}_{m}^{(2)})^{2} \sigma^{'}(\langle \hat{\textbf{w}}_{m}^{(1)}, \textbf{x}_{1} \rangle) \sigma^{'}(\langle \hat{\textbf{w}}_{m}^{(1)}, \textbf{x}_{2} \rangle) \left\langle \textbf{x}_{1}, C_{\boldsymbol{\alpha}} \textbf{x}_{2} \right\rangle \right]},
\end{equation}

\begin{equation}
\label{eq:t2}
\small{\mathcal{K}_{\rm vqc}^{(\boldsymbol{\beta})}(\textbf{x}_{1}, \textbf{x}_{2})  \xrightarrow{M \rightarrow \infty} \mathbb{E}_{\left(\boldsymbol{\theta}_{\rm vqc}, \boldsymbol{\theta}_{W^{(2)}}\right)} \left[ (\textbf{w}_{m}^{(2)})^{2} \sigma^{'}(\langle \hat{\textbf{w}}_{m}^{(1)}, \textbf{x}_{1} \rangle) \sigma^{'}(\langle \hat{\textbf{w}}_{m}^{(1)}, \textbf{x}_{2} \rangle) \left\langle \textbf{x}_{1}, C_{\boldsymbol{\beta}} \textbf{x}_{2} \right\rangle \right]},
\end{equation}

\begin{equation}
\label{eq:t3}
\small{\mathcal{K}_{\rm vqc}^{(\boldsymbol{\gamma})}(\textbf{x}_{1}, \textbf{x}_{2})  \xrightarrow{M \rightarrow \infty} \mathbb{E}_{\left(\boldsymbol{\theta}_{\rm vqc}, \boldsymbol{\theta}_{W^{(2)}}\right)} \left[ (\textbf{w}_{m}^{(2)})^{2} \sigma^{'}(\langle \hat{\textbf{w}}_{m}^{(1)}, \textbf{x}_{1} \rangle) \sigma^{'}(\langle \hat{\textbf{w}}_{m}^{(1)}, \textbf{x}_{2} \rangle) \left\langle \textbf{x}_{1}, C_{\boldsymbol{\gamma}} \textbf{x}_{2} \right\rangle \right]},
\end{equation}
\vspace{1mm}
\begin{equation}
\begin{split}
\mathcal{K}_{W^{(2)}}(\textbf{x}_{1}, \textbf{x}_{2})  \xrightarrow{M \rightarrow \infty}  \mathbb{E}_{\boldsymbol{\theta}_{\rm vqc}}\left[ \sigma^{'}\left( \langle \hat{\textbf{w}}_{m}^{(1)}, \textbf{x}_{1} \rangle \right) \sigma^{'}\left( \langle \hat{\textbf{w}}_{m}^{(1)}, \textbf{x}_{2} \rangle \right) \right]. 
\end{split}
\end{equation}

Thus, the over-parameterized regime (large-width) ensures the NTK matrix $\mathcal{K}_{\rm vm}$ remains nearly constant throughout training. Furthermore, since classical kernels often have better-conditioned NTKs~\cite{lee2019wide}, introducing the classical component can boost the lowest eigenvalue: $\lambda_{\min}(\mathcal{K}_{\rm vm}) \gg \lambda_{\min}(\mathcal{K}_{\rm vqc})$. 

Next, we prove the upper bound on the optimization error based on the derived NTK $\mathcal{K}_{\rm vm}$ and $\lambda_{\min}(\mathcal{K}_{\rm vm})$. Given the set of training data $\{(\textbf{x}_{1}, y_{1}), (\textbf{x}_{2}, y_{2}), ..., (\textbf{x}_{N}, y_{N})\}$, we aim to minimize the cross-entropy loss: 
\begin{equation}
\hat{\mathcal{R}}(f_{\boldsymbol{\theta}}) = -\frac{1}{N} \sum\limits_{n=1}^{N} y_{n} \log\left(	 \sigma_{c}(f_{\boldsymbol{\theta}}(\textbf{x}_{n})) \right), 
\end{equation}
where $\sigma_{c}(\cdot)$ denotes the softmax probabilities. 

Since the parameter update follows gradient flow dynamics: 
\begin{equation}
\frac{d\boldsymbol{\theta}}{dt} = -\nabla_{\boldsymbol{\theta}}\mathcal{\hat{R}}(f_{\boldsymbol{\theta}}), 
\end{equation}
we have: 
\begin{equation}
\frac{d}{dt}f_{\boldsymbol{\theta}}(\textbf{X}) = \frac{d f_{\boldsymbol{\theta}}}{d\boldsymbol{\theta}}\frac{d\boldsymbol{\theta}}{dt} = -\mathcal{K}_{\rm vm}\cdot \left( \sigma_{c}(f_{\boldsymbol{\theta}}(\textbf{X})) - \textbf{y} \right),
\end{equation}
where we define the data matrix $\textbf{X} = [\textbf{x}_{1} \textbf{x}_{2} ... \textbf{x}_{N}]$ and the label vector $\textbf{y} = [y_1 y_2 ... y_{N}]^{\top}$.

Furthermore, due to the nonlinearity of the activation function $\sigma_{\rm c}(\cdot)$, we further simplify it by expanding the softmax around the initial predictions. Typically, this linearization approximation is: 
\begin{equation}
\sigma_{c}(f_{\boldsymbol{\theta}_{t}}(\textbf{X})) \approx \sigma_{c}(f_{\boldsymbol{\theta}_{0}}(\textbf{X})) + \nabla_{\boldsymbol{\theta}}f_{\boldsymbol{\theta}_{0}}(\textbf{X})^{\top} \left( f_{\boldsymbol{\theta}_{t}}(\textbf{X}) - f_{\boldsymbol{\theta}_{0}}(\textbf{X}) \right).
\end{equation}

Now we have a linear, time-invariant ordinary differential equation (ODE): 
\begin{equation}
\frac{d}{dt}f_{\boldsymbol{\theta}_{t}}(\textbf{X}) \approx -\mathcal{K}_{\rm vm} \cdot \left( [\sigma_{\rm c}(f_{\boldsymbol{\theta}_{0}}(\textbf{X})) - \textbf{y}] +  \nabla_{\boldsymbol{\theta}}f_{\boldsymbol{\theta}_{0}}(\textbf{X})^{\top}(f_{\boldsymbol{\theta}_{t}}(\textbf{X}) - f_{\boldsymbol{\theta}_{0}}(\textbf{X})) \right).
\end{equation}

Solving the above ODE, we obtain a closed form of $f_{\boldsymbol{\theta}_{t}}(\textbf{X})$ as: 
\begin{equation}
f_{\boldsymbol{\theta}_{t}}(\textbf{X}) = f_{\boldsymbol{\theta}_{0}}(\textbf{X}) - \nabla_{\boldsymbol{\theta}}f_{\boldsymbol{\theta}_{0}}(\textbf{X})^{-1}\left(I - e^{-\mathcal{K}_{\rm vm} \cdot \nabla_{\boldsymbol{\theta}}f_{\boldsymbol{\theta}_{0}}(\textbf{X}) t} \right)(\sigma_{c}(f_{\boldsymbol{\theta}_{0}}(\textbf{X})) - \textbf{y}). 
\end{equation}

To minimize $\tilde{f}_{\boldsymbol{\theta}_{t}}(\textbf{X})$, we obtain the optimal $\boldsymbol{\theta}^{*}$ such that 
\begin{equation}
f_{\boldsymbol{\theta}^{*}}(\textbf{X}) = f_{\boldsymbol{\theta}_{\infty}}(\textbf{X}) = f_{\boldsymbol{\theta}_{0}}(\textbf{X}) - (\nabla_{\boldsymbol{\theta}}f_{\boldsymbol{\theta}_{0}}(\textbf{X}))^{-1}(\sigma_{c}(f_{\boldsymbol{\theta}_{0}}(\textbf{X})) - \textbf{y}). 
\end{equation}

As for the optimization error $\epsilon_{\rm opt}$, at epoch t, we set $\boldsymbol{\theta}_{t} = \boldsymbol{\hat{\theta}}$ and further derive that:
\begin{equation}
\begin{split}
\mathcal{\hat{R}}(f_{\boldsymbol{\hat{\theta}}}) -   \mathcal{\hat{R}}(f_{\boldsymbol{\theta}^{*}}) &\le L_{\rm ce} \lVert	f_{\boldsymbol{\theta}_{t}} - f_{\boldsymbol{\theta}^{*}} \rVert_{2}	\\
&= L_{\rm ce} \left\lVert (\nabla_{\boldsymbol{\theta}}f_{\boldsymbol{\theta}_{0}}(\textbf{X}))^{-1} e^{-\mathcal{K}_{\rm vm}\cdot \nabla_{\boldsymbol{\theta}}f_{\boldsymbol{\theta}_{0}}(\textbf{X}) t} (\sigma_{c}(f_{\boldsymbol{\theta}_{0}}(\textbf{X})) - \textbf{y}) \right\rVert_{2} \\
&\le L_{\rm ce} \left\lVert	 (\nabla_{\boldsymbol{\theta}}f_{\boldsymbol{\theta}_{0}}(\textbf{X}))^{-1}  \right\rVert_{2} \left\lVert	e^{-\mathcal{K}_{\rm vm} \cdot \nabla_{\boldsymbol{\theta}}f_{\boldsymbol{\theta}_{0}}(\textbf{X}) t} \right\rVert_{2} \left\lVert \sigma_{c}(f_{\boldsymbol{\theta}_{0}}(\textbf{X})) - \textbf{y} \right\rVert_{2}. 
\end{split}
\end{equation}

Since both $\mathcal{K}_{\rm vm}$ and $\nabla_{\boldsymbol{\theta}}f_{\boldsymbol{\theta}_{0}}(\textbf{X})$ are positive semi-definite matrices, using the eigendecomposition techniques, we have:
\begin{equation}
\left\lVert (\nabla_{\boldsymbol{\theta}}f_{\boldsymbol{\theta}_{0}}(\textbf{X}))^{-1}  \right\rVert_{2} = \frac{1}{\lambda_{\rm min}(\nabla_{\boldsymbol{\theta}}f_{\boldsymbol{\theta}_{0}}(\textbf{X}))}, 
\end{equation}

\begin{equation}
\left\lVert e^{-\mathcal{K}_{\rm vm}\cdot \nabla_{\boldsymbol{\theta}}f_{\boldsymbol{\theta}_{0}}(\textbf{X}) t} \right\rVert_{2} = e^{-\lambda_{\min}\left(\mathcal{K}_{\rm vm}\cdot \nabla_{\boldsymbol{\theta}}f_{\boldsymbol{\theta}_{0}}(\textbf{X})\right) t}. 
\end{equation}

Thus, we verify the upper bound on the optimization error as: 
\begin{equation}
\epsilon_{\rm opt} = \sup\limits_{\boldsymbol{\hat{\theta}} \in \Theta}\left( \mathcal{\hat{R}}(f_{\boldsymbol{\hat{\theta}}}) -   \mathcal{\hat{R}}(f_{\boldsymbol{\theta}^{*}}) \right) \le L_{\rm ce} \cdot \frac{\lVert	 \sigma_{c}(f_{\boldsymbol{\theta}_{0}}(\textbf{X})) - \textbf{y} \rVert_{2}}{\lambda_{\rm min} (\nabla_{\boldsymbol{\theta}}f_{\boldsymbol{\theta}_{0}}(\textbf{X}))}\cdot e^{-\lambda_{\min}\left(\mathcal{K}_{\rm vm} \cdot \nabla_{\boldsymbol{\theta}}f_{\boldsymbol{\theta}_{0}}(\textbf{X})\right) t}. 
\end{equation}

In practice, since the term $\nabla_{\boldsymbol{\theta}}f_{\boldsymbol{\theta}_{0}}(\textbf{X})$ is typically bounded and stable, the primary determinant of convergence rate corresponds to the minimum eigenvalue $\lambda_{\rm min}(\mathcal{K}_{\rm vm})$. Thus, we simplify the above term as:
\begin{equation}
\epsilon_{\rm opt} = \sup\limits_{\boldsymbol{\hat{\theta}} \in \Theta}\left( \mathcal{\hat{R}}(f_{\boldsymbol{\hat{\theta}}}) -   \mathcal{\hat{R}}(f_{\boldsymbol{\theta}^{*}}) \right) \le C_{0} e^{-\lambda_{\min}\left(\mathcal{K}_{\rm vm} \right) t}, 
\end{equation}
where the constant $C_{0} =  \frac{L_{\rm ce}\cdot \lVert \sigma_{c}(f_{\boldsymbol{\theta}_{0}}(\textbf{X})) - \textbf{y} \rVert_{2}}{\lambda_{\rm min} (\nabla_{\boldsymbol{\theta}}f_{\boldsymbol{\theta}_{0}}(\textbf{X}))}$. 

\vspace{1mm}

This analysis shows that the VQC-MLPNet benefits from classical trainability and quantum expressivity. The hybrid NTK structure stabilizes training by providing a favorable eigenvalue spectrum, ensuring effective convergence under practical optimization schemes.

\section{DATA AVAILABILITY}
The dataset used in our experiments on quantum dot classification can be downloaded from the website: https://gitlab.com/QMAI/mlqe$\_$2023$\_$edx. The dataset for TFBS predictions can be accessed via https://www.ebi.ac.uk/interpro/entry/InterPro/IPR029823.

\section{CODE AVAILABILITY}
Our codes for VQC-MLPNet and other VQC models are available on the website: https://github.com/jqi41/VQC-MLPNet.  

\section{References}

\bibliographystyle{IEEEbib}
\bibliography{sn-bibliography}

\section{FUNDING DECLARATION}
This work is partly funded by the Hong Kong Research Impact Fund (R6011-23).

\section{COMPETING INTERESTS}
The authors declare no Competing Financial or Non-Financial Interests.

\section{AUTHOR CONTRIBUTIONS}
Jun Qi and Chao-Han Yang conceived the project. Jun Qi and Min-Hsiu Hsieh completed the theoretical analysis. Jun Qi, Chao-Han, and Pin-Yu Chen designed the experimental work. Min-Hsiu Hsieh and Pin-Yu Chen provided high-level advice on the paperwork pipeline, and Jun Qi wrote the manuscript. 

\end{document}